\documentclass[final,5p,times,twocolumn]{elsarticle}

\usepackage{graphicx}
\usepackage{graphicx}% Include figure files
\usepackage{dcolumn}% Align table columns on decimal point
\usepackage{bm}% bold math
\usepackage{subfigure}% sub figure
\usepackage{amsmath}
\usepackage{graphicx}
\usepackage{amssymb}

\def\gsim{\;\lower4pt\hbox{${\buildrel\displaystyle >\over\sim}$}\;}
\def\lsim{\;\lower4pt\hbox{${\buildrel\displaystyle <\over\sim}$}\;}
\def\grls{\;\lower4pt\hbox{${\buildrel\displaystyle >\over <}$}\;}

\journal{New Astronomy}

\begin{document}

\begin{frontmatter}

\title{General Polytropic Magnetofluid under Self-Gravity:
Voids and Shocks}

\author[1,2,3]{Yu-Qing Lou\corref{Yu-Qing Lou}}\ead{
louyq@tsinghua.edu.cn, lou@oddjob.uchicago.edu} \cortext[Yu-Qing
Lou]{Corresponding author.}
\author[1]{  Ren-Yu Hu}\ead{hu-ry07@mails.tsinghua.edu.cn}

\address[1]{Physics Department and Tsinghua Center for
Astrophysics (THCA), Tsinghua University, Beijing 100084, China}
\address[2]{Department of Astronomy and Astrophysics, The University
of Chicago, 5640 S. Ellis Avenue, Chicago, IL 60637 USA}
\address[3]{National Astronomical Observatories, Chinese Academy
of Sciences, A20, Datun Road, Beijing 100012, China }

\date{Received: date / Accepted: date}

\begin{abstract}
We study the self-similar magnetohydrodynamics (MHD) of a
quasi-spherical expanding void (viz. cavity or bubble) in the
centre of a self-gravitating gas sphere with a general polytropic
equation of state. We show various analytic asymptotic solutions
near the void boundary in different parameter regimes and obtain
the corresponding void solutions by extensive numerical
explorations. We find novel void solutions of zero density on the
void boundary. These new void solutions exist only in a general
polytropic gas and feature shell-type density profiles. These void
solutions, if not encountering the magnetosonic critical curve
(MCC), generally approach the asymptotic expansion solution far
from the central void with a velocity proportional to radial
distance.
%(Removed by Hu on April 21 in response to the referee)
%We examine free-expansion solutions with the parameter $q<2/3$,
%Einstein-de Sitter expansion solutions with $q=2/3$, and
%thermal-expansion solutions with $q>2/3$, where parameter $q\equiv
%2(n+\gamma-2)/(3n-2)$, $n$ is a key scaling exponent in the
%self-similar transformation and $\gamma$ is the polytropic index.
We identify and examine free-expansion solutions, Einstein-de
Sitter expansion solutions, and thermal-expansion solutions in
three different parameter regimes. Under certain conditions, void
solutions may cross the MCC either smoothly or by MHD shocks, and
then merge into asymptotic solutions with finite velocity and
density far from the centre. Our general polytropic MHD void
solutions provide physical insight for void evolution, and may
have astrophysical applications such as massive star collapses and
explosions, shell-type supernova remnants and hot bubbles in the
interstellar and intergalactic media, and planetary nebulae.
\end{abstract}


\end{frontmatter}

\section[]{Introduction}\label{intro}
\renewcommand{\topfraction}{0.98}

Supernova explosions, planetary nebulae and stellar winds from
massive stars are believed to be the main sources of creating voids
(i.e. bubbles, cavities) in the interstellar medium (ISM) (e.g.
Ferri\`ere 1998, 2001 and extensive references therein). In the
local ISM, the remnant of a typical isolated supernova grows for
$\sim$ 1.5 Myr and reaches a maximum radius of $\sim 50$ pc.
%There are two major morphological types of supernova remnants (SNRs).
%Filled-centre remnants are amorphous, centrally brightened nebulae
%having a flat radio spectrum, thought to be powered by the central
%pulsar (the Crab Nebula is a well-known example).
The shell-type supernova remnants (SNRs) appear quasi-spherical and
the masses within them have been swept up by ejecta from supernovae.
An example of such structure is a void region towards the Lupus dark
cloud complex, which was revealed through observations of 100 $\mu$m
emissions (e.g. Gahm et al. 1990; Franco 2002)
%\citep[e.g.][]{b54,b57},
soft X-ray (e.g. Riegler et al. 1980),
%\citep[e.g.][]{b55},
and 21 cm HI line (e.g. Colomb et al. 1984).
%\citep[e.g.][]{b56}.
Neutral hydrogen (HI) voids have also been found in filled-centre
SNRs (e.g. Wallace et al. 1994).
%\citep[e.g.][]{b70}.

Voids may also emerge long before we actually observe SNRs.
According to the neutrino-driven mechanism of type-II and type-Ibc
supernova explosions (e.g. Janka \& Hillebrant 1989; Janka \&
M\"uller 1995, 1996),
%\citep[e.g.][]{Janka3,Janka1,Janka2},
within the first second of a type II supernova, the intense
neutrino flux generated by the central core bounce heats the
surrounding stellar mass and pushes the stellar material outwards.
Subsequently, a rebound shock emerges and propagates outwards
(e.g. Lou \& Wang 2006, 2007; Hu \& Lou 2009).
%\citep[e.g.][]{b11,b12}.
%The convection inside the neutrino-sphere and beneath
%the shock influences the eventual evolution.
After several hundred milliseconds, the neutrino-sphere decouples
from the gas, and may leave behind a cavity around the centre
during a supernova. After this decoupling, the exploding star with
a central cavity continues to expand and the central cavity
eventually evolves into a hot bubble in the ISM.

Massive stars can also create voids in the ISM through
photoionization heating and stellar winds (e.g. Castor et al. 1975;
Weaver et al. 1977; McKee et al. 1984).
%\citep[e.g.][]{b59,b60,b61}.
Likewise, fast winds from central compact hot white dwarfs may
generate expanding cavities in planetary nebulae (e.g. Lou \& Zhai
2009). HI voids and shells have been found and observed by radio
observations around several Galactic Wolf-Rayet (WR) stars --
massive stars undergoing significant mass losses (e.g. Cappa \&
Miemela 1984; Cappa et al. 1986, 1988; Dubner et al. 1990; Niemela
\& Cappa 1991; Arnal \& Mirabel 1991; Arnal 1992).
%\citep[e.g.][]{b62,b63,b64,b65,b66,b67,b68}.
%It should be noted that the cavity in the vicinity of a massive
%star is not totally empty; a massive star resides at the centre
%and there are rarefied stellar wind materials. By ignoring the
%central stellar mass and the mass of wind materials, the
%evolution of such stellar systems can be idealized as a void.
On larger scales, observations show that voids also exist in neutral
hydrogen discs of spiral galaxies (e.g. Crosthwaite \& Turner 2000).
%\citep[e.g.][]{b69}.
Recently, central cavities of $\sim 200$ kpc diameter and
large-scale shock fronts have been revealed by {\it Chandra} X-ray
observations in the galaxy cluster MS0735.6+7421 (e.g. McNamara et
al. 2005).
%\citep[e.g.][]{McNamara}.

The dynamic evolution of voids in the ISM still lacks a systematic
theoretical exploration.
%\citet{DW}
Dyson \& Williams (1997) provided qualitative description on the gas
dynamic effects of massive stars on the ISM. Chevalier (1997)
%\citet{Chevalier1997}
studied the expansion of a photoionized stellar wind in late
stages of stellar evolution (e.g. supernovae and planetary
nebulae; see also Meyer 1997). From the centre, a stellar system
consists of a hot bubble of shocked fast wind, a region of shocked
and photoionized wind, and an outer region of slow wind.
%\citet{Chevalier1997}
Chevalier (1997) also employed the isothermal self-similar
transformation as Shu (1977) but without gravity to model the
self-similar dynamic evolution of an outer slow wind. Physically,
the fast hot wind bubble resembles the concept of a central void
of this paper. Hu \& Lou (2008a) presented self-similar void
solutions to model ``champagne flows'' of H II regions after the
nascence of a massive protostar in a conventional polytropic gas.
Such voids embedded in nebulae can be created and sustained by
fast stellar winds and photoionization heating. Here, we formulate
an MHD problem with a general polytropic gas under self-gravity.

The system of interest is a general polytropic magnetofluid with a
quasi-spherical symmetry under self-gravity, thermal pressure
gradient force and magnetic Lorentz force. We further find that
MHD shocks are indispensable to establish sensible global
solutions, for example with the asymptotic velocity at large radii
tending to zero. Magnetic field can be extremely important in many
astrophysical processes on different scales and in particular, for
star formation activities at various stages (e.g. Shu et al. 1987;
Myers 1998). The Crab Nebula is observed to be supported by the
magnetized pulsar wind (e.g. Lou 1993; Wolf et al. 2003). When
rotation is sufficiently slow in an astrophysical system, the
overall geometry may remain quasi-spherical and the
quasi-spherical random-field approximation (e.g. Zel'dovich \&
Novikov 1971) can be applicable. Chiueh \& Chou (1994) discussed
the gravitational collapse of an isothermal magnetized gas cloud
by including the magnetic pressure force from a randomly
distributed magnetic field. Recently, we have provided detailed
analyses by assuming a random transverse magnetic field with the
consideration of both magnetic pressure and tension forces (Yu \&
Lou 2005; Yu et al. 2006; Lou \& Wang 2007; Wang \& Lou 2007,
2008). We presume that the non-spherical flows as a result of the
magnetic tension force may be neglected as compared to the
large-scale mean radial bulk motion of gas. The key point is that
the (self-)gravity is strong enough to hold on the entire gas mass
and induce core collapse, or the driving force is strong enough to
generate a quasi-spherical void. Therefore on large scales a
completely random magnetic field contributes to the dynamics in
the form of the average magnetic pressure gradient force and the
average magnetic tension force in the radial direction. This
approximation was discussed in more details by Wang \& Lou (2007).

The general equation of state is $p=\kappa(r,\ t)\rho^{\gamma}$
where $p$ is the thermal gas pressure, $\rho$ is the mass density,
$\gamma$ is the polytropic index and $\kappa$ is a proportional
coefficient dependent on both radius $r$ and time $t$. For a
global constant $\kappa$, the equation of state is that of a
conventional polytropic gas (e.g. Suto \& Silk 1988; Lou \& Gao
2006; Lou \& Wang 2006, 2007; Hu \& Lou 2008a). By setting
$\gamma=1$ and $\kappa$ as a global constant, a conventional
polytropic gas then becomes an isothermal gas. In case of
$\gamma>1.2$, novel quasi-static asymptotic solutions for a
polytropic gas exist in approach to the system centre (see Lou \&
Wang 2006). As the specific enthalpy is $p/(\gamma-1)$, we thus
require $\gamma\geq 1$ to ensure a positive specific enthalpy. A
general polytropic gas features the conservation of specific
entropy along streamlines. The conventional polytropic case is a
only special case with constant specific entropy everywhere at all
times. A general polytropic model with random magnetic field is
the most general model of a polytropic magnetofluid of
quasi-spherical symmetry under self-gravity (Wang \& Lou 2008;
Jiang \& Lou 2009).

Our self-similar transformation employs a dimensionless
independent similarity variable $x$ defined as a combination of
radius $r$ and time $t$ such that $x=r/(k^{1/2}t^n)$ where $k$ is
the so-called `sound parameter' to make $x$ dimensionless and $n$
is a key scaling index. Theoretically, our void solutions are
those solutions whose enclosed mass is zero within a certain
radius denoted as $x^*$. This radius expands with time in a
self-similar manner, i.e. $r^*=k^{1/2}x^*t^n$. By this expression,
the physical meaning of the self-similar scaling index parameter
$n$ is evident.
%We show presently the scaling for the radius of void boundary
%$r^* \propto t^n$, where $n$ is a scaling exponent related to
%the polytropic index $\gamma$ and the parameter $q$.
The expansion speed of the void boundary is $u^*\propto
t^{(n-1)}$; therefore for $n>1$, a void expands faster and faster
(i.e. acceleration), while for $n<1$, a void expands slower and
slower (i.e. deceleration); and for $n=1$, a void expands at a
constant speed. The void boundary may also be regarded as an
idealization of a contact discontinuity between a faster wind and
a slower winds (e.g. Chevalier 1997; Lou \& Zhai 2009).

The case of $\gamma=4/3$ corresponds to a relativistically hot gas
that deserves a special attention. Homologous core collapse for a
relativistically hot gas was studied by Goldreich \& Weber (1980)
%\citet{b29}
and the behaviour of such system has been treated by Yahil (1983)
%\citet{b30}
as a limit of $\gamma\rightarrow (4/3)^+$. Recently, Lou \& Cao
(2008)
%\citet{b5}
presented an illustrative example of void in such a system. In
this paper, we study voids for a magnetized Newtonian gas (i.e.
$\gamma\neq 4/3$ in general), as well as voids in a
relativistically hot fluid (i.e. $\gamma=4/3$), and we offer
several concrete examples.
%Our model is generally suitable for astrophysical systems, such as
%stars, planetary nebulae, and dust or molecular clouds in galaxies.
%It is clear that our expanding void solutions bear conceptual
%relevance to shell-type SNRs.

This paper is structured as follows. Section 1 is an introduction
to provide background information and the motivation of this
investigation; Section 2 describes first the formulation of a
general polytropic magnetofluid under quasi-spherical symmetry,
and secondly analytic asymptotic solution behaviours near the void
boundary in various parameter regimes, and thirdly various
asymptotic MHD solutions that are useful in constructing global
semi-complete solutions (i.e. solutions that are valid in the
range $0<x< +\infty$); Section 3 describes and discusses
properties of void solutions with different parameters, in
contexts of hydrodynamics and MHD, and presents a few examples;
Section 4 gives examples of astrophysical applications of such
void solutions. Finally, Section 5 contains conclusions and
discussion. Mathematical derivations are included in an appendix.

%{\bf Stop reading here}

\section[]{Self-Similar MHD with Quasi-Spherical Symmetry}

\subsection[]{Formulation of a Nonlinear MHD Problem}

The MHD evolution of a general polytropic gas of quasi-spherical
symmetry and under self-gravity can be described by a set of
nonlinear MHD partial differential equations (PDEs) in spherical
polar coordinates $(r,\ \theta,\ \phi)$, namely
\begin{equation}
\frac{\partial \rho}{\partial
t}+\frac{1}{r^2}\frac{\partial}{\partial r}(r^2\rho u)=0\ ,
\label{equ1}
\end{equation}
%\begin{equation}
%\frac{\partial M}{\partial t}+u\frac{\partial M}{\partial r}=0\ ,
%\label{equ2}
%\end{equation}
\begin{equation}
\frac{\partial M}{\partial r}=4\pi r^2\rho\ ,\label{equ3}
\end{equation}
\begin{equation}
\rho\frac{\partial u}{\partial t} +\rho u\frac{\partial u}{\partial
r}=-\frac{\partial p}{\partial r}-\frac{GM\rho
}{r^2}-\frac{\partial}{\partial r}
\frac{<B_t^2>}{8\pi}-\frac{<B_t^2>}{4\pi r}\ ,\label{equ4}
\end{equation}
where $\rho(r,\ t)$ is the mass density, $u(r,\ t)$ is the bulk
radial gas flow speed, $M(r,\ t)$ is the enclosed mass within
radius $r$ at time $t$, $p$ is the thermal gas pressure,
$G=6.67\times 10^{-8}$ dyne cm$^2$ g$^{-2}$ is the gravitational
constant, and $<B_t^2>$ is the ensemble average of a random
transverse magnetic field squared (i.e. proportional to the
magnetic energy density). Equations (\ref{equ1}) and (\ref{equ3})
represent mass conservation, leading to ${\partial M}/{\partial
t}+u{\partial M}/{\partial r}=0$. We assume a random magnetic
field mainly in transverse directions, and the magnetic force
perpendicular to magnetic field lines
% under the `frozen-in' condition,
directs to the radial direction and appears in the radial momentum
equation (\ref{equ4}) as the magnetic pressure and tension terms.
%The Poisson
%equation relating the gravitational potential $\Phi(r,\ t)$ and the
%mass density $\rho(r,\ t)$ is automatically satisfied.
Together with magnetic induction equation
\begin{equation}
\bigg(\frac{\partial}{\partial t}+u \frac{\partial}{\partial
r}\bigg)(r^2 <B_t^2>)+2r^2<B_t^2>\frac{\partial u}{\partial r}=0\
 \label{equ5}
\end{equation}
and the specific entropy conservation along streamlines
\begin{equation}
\bigg(\frac{\partial}{\partial t}+u\frac{\partial}{\partial
r}\bigg)\bigg(\ln\frac{p}{\rho^\gamma}\bigg)=0\ \label{equ6}
\end{equation}
with $\gamma$ being the polytropic index, we complete the model
formulation for a general polytropic MHD with a quasi-spherical
symmetry (Wang \& Lou 2007, 2008).
%This set of MHD PDEs is the same as that of \citep{b3,b4}.
%Equation of state (\ref{equ6}) differs from previous
%formulations of MHD gas under self-gravity in that
%conservation of `specific entropy' along streamlines allows
%the `specific entropy' to vary in radius $r$ and time $t$.

%{\bf Stop reading here}

In this paper, we consider self-similar solutions which form an
important subclass of nonlinear MHD PDEs. In order to reduce these
nonlinear MHD PDEs to self-similar ordinary differential equations
(ODEs), we introduce the following self-similar transformation as
Wang \& Lou (2008), namely
\begin{eqnarray}
\!\!\!\!\!\!\!\!\!\! &&\!\!\!\!\!\!\!\!\! r=k^{1/2}t^n x,\qquad
u=k^{1/2} t^{n-1} v\ ,\qquad\quad\rho=\frac{\alpha}{4\pi G
t^2}\ ,\qquad\nonumber\\
\nonumber\\
&&\!\!\!\!\!\!\!\!\!\!\!\!\!\!\!\!\!  p=\frac{k t^{2n-4}}{4\pi
G}\beta\ ,\quad M=\frac{k^{3/2} t^{3n-2} m}{(3n-2)G}\ ,\quad
<B_t^2>=\frac{k t^{2n-4}w}{G},\label{equ7}
\end{eqnarray}
where $v(x)$, $\alpha(x)$, $\beta(x)$, $m(x)$, $w(x)$ are
dimensionless reduced variables of $x$ only. We refer to $v(x)$,
$\alpha(x)$, $\beta(x)$, $m(x)$ and $w(x)$ as the reduced radial
speed, mass density, thermal pressure, enclosed mass and magnetic
energy density, respectively. Self-similar transformation
(\ref{equ7}) is identical with that of Wang \& Lou (2007, 2008).
By substituting self-similar transformation (\ref{equ7}) into
equations (\ref{equ1})$-$(\ref{equ6}), we obtain several valuable
integrals
\begin{equation}
w=h\alpha^2 x^2\ ,\label{equ9}
\end{equation}
\begin{equation}
\beta=C\alpha^\gamma m^q\ ,\label{equ10}
\end{equation}
\begin{equation}
m=\alpha x^2(nx-v)\ .\label{equ8}
\end{equation}
Equation (\ref{equ9}) corresponds to the frozen-in condition for
magnetic field in the ideal MHD approximation, where $h\equiv
<B_t^2>/(16\pi^2 G\rho^2r^2)$ is a dimensionless magnetic
parameter representing the average strength of a random transverse
magnetic field.

Equation (\ref{equ10}) is the self-similar form of the specific
entropy conservation along streamlines, where the exponent
parameter $q\equiv 2(n+\gamma-2)/(3n-2)$ and thus
$\gamma=2-n+(3n-2)q/2$, and $C$ is an arbitrary coefficient from
integration. For $q=0$, the flow system involves a conventional
polytropic gas with a constant specific entropy everywhere in
space and at all times; for $q>0$, the specific entropy increases
from inside (smaller $x$) to outside (larger $x$); and $q=2/3$
leads to $\gamma=4/3$ for a relativistically hot gas (e.g. a
photon gas, a neutrino gas or an extremely high temperature
electron gas) with an arbitrary $c$.
%For $n=2/3$, self-similar transformation (\ref{equ7}) is no longer
%applicable. Lou \& Cao (2008) developed a new self-similar
%transformation to deal with such situation. The cases with $n\neq
%2/3,\ q=2/3$ can be treated with our self-similar transformation.
Actually we may set $C=1$ in all cases with $q\neq 2/3$ without
loss of generality, because an adjustment of sound parameter $k$
in self-similar transformation (\ref{equ7}) to $C^{1/(1-3q/2)}k$
would make $C$ disappear.

Equation (\ref{equ8}) requires both $3n-2>0$ and $nx-v>0$ for a
positive enclosed mass $M(r,\ t)$. When $nx-v=0$ at a certain
$x^*$, the enclosed mass vanishes by equation (\ref{equ8}); this
is referred to as a void with $x^*$ being the void boundary in a
self-similar expansion. Accordingly, the reduced radial velocity
on the boundary is given by $v^*=nx^*$. The condition $nx-v=0$ on
the void boundary implies that the self-similar expansion speed of
the void boundary $dr^*/dt$ is equal to the radial flow velocity
on the void boundary $u(r^*,t)$. This is regarded as the physical
condition for a contact discontinuity between the outer slower
stellar wind and the inner faster wind driving a hot bubble by
Chevalier (1997) for $q=0$, $n=1$, $\gamma=1$ and without gravity.
Lou \& Zhai (2009) considered a gas dynamic model for planetary
nebulae with contact discontinuities for an isothermal
self-gravitating gas. From now on, we denote variables on the void
boundary by a superscript asterisk $^*$.

Combining all equations above, we obtain coupled nonlinear MHD
ODEs for the two first derivatives $\alpha'$ and $v'$ in the
following compact forms of
\begin{equation}
%\!\!\!\!\!\!\!
{\cal X}(x,\alpha,v) \alpha'={\cal A}(x,\alpha,v)\ ,\qquad
{\cal
X}(x, \alpha, v)v'={\cal V}(x,\alpha, v)\ ,\label{equ11}
\end{equation}
where the three functional coefficients ${\cal X}$, ${\cal A}$ and
${\cal V}$ are explicitly defined by
\begin{eqnarray}
\!\!\!\!\!\!\!\!&&\!\!\!\!\!\!\!\!\! {\cal X}(x,\ \alpha,\ v)\equiv
C
\bigg[2-n+\frac{(3n-2)}{2}q\bigg]\nonumber\\
&&\quad\times\alpha^{1-n+3nq/2} x^{2q} (nx-v)^q+h\alpha
x^2-(nx-v)^2\ ,\nonumber\end{eqnarray}
\begin{eqnarray}
\!\!\!\!\!\!\!\!&&\!\!\!\!\!\!\!\!\! {\cal A}(x,\ \alpha,\ v)\equiv
2\frac{x-v}{x}\alpha \big[Cq\alpha^{1-n+3nq/2}
x^{2q}(nx-v)^{q-1}\nonumber\\
&&\quad +(nx-v)\big]-\alpha\bigg[(n-1)v+\frac{(nx-v)}{(3n-2)}\alpha
+2h\alpha x
\nonumber\\
&&\quad +Cq\alpha^{1-n+3nq/2}x^{2q-1} (nx-v)^{q-1}(3nx-2v)\bigg]\ ,
\nonumber\end{eqnarray}
\begin{eqnarray}
\!\!\!\!\!\!\!\!&&\!\!\!\!\!\!\!\!\! {\cal V}(x,\ \alpha,\ v)\equiv
2\frac{(x-v)}{x}\alpha\bigg[C
\bigg(2-n+\frac{3n}{2}q\bigg)\nonumber\\
&&\quad\times\alpha^{-n+3nq/2}x^{2q}
(nx-v)^q+hx^2\bigg]\nonumber\\
&&\quad-(nx-v)\bigg[(n-1)v+\frac{(nx-v)}{(3n-2)}\alpha +2h\alpha x
\nonumber\\
&&+Cq\alpha^{1-n+3nq/2}x^{2q-1}(nx-v)^{q-1}(3nx-2v)\bigg]\ .
\label{equ12}
\end{eqnarray}
%With proper boundary and initial asymptotic conditions, coupled
%nonlinear MHD ODEs (\ref{equ11}) and (\ref{equ12}) can be
%numerically integrated using the standard fourth-order Runge-Kutta
%method (e.g. Press et al. 1986).
The formulation above is largely the same as Wang \& Lou (2008),
except for an additional free parameter $C$ in cases with $q=2/3$
(i.e. $\gamma=4/3$). Wang \& Lou (2008) also provides procedures
to determine magnetosonic critical curve (MCC), eigensolutions
across the MCC and MHD shock jump conditions across the
magnetosonic singular surface (see Appendix \ref{MHDshock}).
%We note that for a certain set of $(x,\ \alpha,\ v)$, functional
%${\cal X}$ vanishes, corresponding to the so-called magnetosonic
%singular surface (MSS). Self-similar solutions cannot cross this
%singular surface unless they cross it via an MHD shock, or they
%satisfy the conditions that functionals ${\cal A}$ and ${\cal V}$
%vanish simultaneously. There are only two independent conditions
%among ${\cal X}=0$, ${\cal A}=0$ and ${\cal V}=0$ and they determine
%a magnetosonic critical curve (MCC). The MCC can be readily computed
%by the numerical procedure of Wang \& Lou (2008).
%%\citet{b4}.
%We can further numerically determine the eigensolutions for the
%first derivatives $\alpha^{\prime}$ and $v^{\prime}$ across such a
%MCC.

%\textbf{
The MCC for $q=2/3$ and thus $\gamma=4/3$ is special. Extensive
numerical explorations suggest that the MCC bears the simple form
of $v=\eta x,\ \alpha=$ constant, where $\eta$ is a constant
coefficient dependent on parameters $n,\ h$ and proportional
factor $C$. Substituting this form into equation (\ref{equ12}) and
the MCC conditions ${\cal X}=0$ and ${\cal A}=0$ become
respectively
\begin{eqnarray}
&&\alpha=\frac{(n-\eta)^2}{h+({4}/{3})C(n-\eta)^{2/3}}\ ,\label{MCC231}\\
\nonumber\\
&&\frac{2}{3}C\alpha(n-\eta)^{-1/3}(2-3n)+2(1-\eta)(n-\eta)\nonumber\\
&&
\quad=(n-\eta)\eta+\frac{(n-\eta)}{(3n-2)}\alpha+2h\alpha\
.\label{MCC232}
\end{eqnarray}
Substituting relation (\ref{MCC231}) into equation (\ref{MCC232}) to
eliminate $\alpha$, we immediately derive an expression for the
constant coefficient $\eta$ in terms of $n,\ h$ and $C$. Once $\eta$
is known in relation (\ref{MCC231}), we can compute $\alpha$ value
accordingly. With $h=0$, relations (\ref{MCC231}) and (\ref{MCC232})
give the same solution as discussed in Lou \& Cao (2008)
%\citet{b5}
for a nonmagnetized relativistically hot gas. Here, we extend the
special MCC with $\gamma=4/3$ to a magnetofluid embedded with a
random transverse magnetic field.

\subsection[]{Behaviours of a Polytropic Void Boundary}

We now analyze asymptotic behaviours around the void boundary $x^*$,
and these boundary conditions will be used to construct various
solutions in numerical integrations starting from the void boundary.

The gas pressure should be continuous across the void boundary,
otherwise a shrinkage of the void boundary with diffusions would
be expected. According to equation (\ref{equ10}) and for
$\alpha\neq 0$, inequality $q<0$ on $nx^*-v=0$ leads to a
diverging reduced pressure $\beta$. For $q\geq 0$, it follows
automatically that $\beta=0$ at the void boundary. Therefore to
ensure $\beta^*\equiv\beta(x^*)=0$, we require
$\alpha*\equiv\alpha(x^*)=0$ in cases of $q<0$, or $\alpha^*\geq
0$ in cases of $q\geq 0$. It is favorable to further require the
continuity of mass density, as a discontinuous density would lead
to a local diffusion, in addition to a global self-similar
evolution. Therefore, a solution with $\alpha=0$ at $nx^*-v=0$ is
regarded as a physically sensible one, otherwise the self-similar
solution should be seen as an asymptotic solution valid
in the region sufficiently far from the void boundary.
%The diffusive behaviour caused by a discontinuous mass density
%is beyond the scope of this paper, but we could still approximate
%such a system as a self-similar evolution on large scales.

\subsubsection[]{Hydrodynamic and MHD Cases with
$\alpha^*=0$
\\ \qquad\qquad on the Expanding Void Boundary}

The void boundary obeying $\alpha=0$ and $nx-v=0$ may possibly
become a critical curve with the three functional coefficients on
both sides of equations (\ref{equ11}) being zero.
%{\bf Meaning?}
According to equation (\ref{equ12}), the possible non-zero terms of
the three functional coefficients ${\cal X}$, ${\cal A}$ and ${\cal
V}$ approaching the void boundary $nx-v=0$ and $\alpha=0$ are the
thermal pressure gradient force terms, namely
\begin{eqnarray}
&&{\cal X}\sim C\gamma
x^{2q}\alpha^{\gamma+q-1}(nx-v)^q\ ,\nonumber\\
\nonumber\\
&&{\cal A}\sim C(2-3n)qx^{2q}\alpha^{\gamma+q}(nx-v)^{q-1}\ ,\nonumber\\
\nonumber\\
&&\!\!\!\!\!\!\!\!\!\!\!\!\!\! {\cal V}\sim C[2(1-n)\gamma+(2-3n)q]
%\nonumber\\ &&\qquad\qquad\qquad\qquad\times
x^{2q}\alpha^{\gamma+q-1}(nx-v)^q\ ,\nonumber\\
\label{XAV}
\end{eqnarray}
where we have used the relation $\gamma=2-n+(3n-2)q/2$ .

According to expressions (\ref{XAV}), the parameter regime of
$q\geq 0\ ,\ \gamma+q\geq 1$, except for the special isothermal
case ($q=0,\ \gamma+q=1$), ensures the vanishing of ${\cal X}$ and
${\cal V}$ on the void boundary; when $\alpha\sim (nx-v)$ near the
void boundary as shown by our analysis presently, then ${\cal A}$
also vanishes at the void boundary, and the void boundary indeed
becomes a MCC.

By a local first-order Taylor series expansion, we obtain from
equations (\ref{equ11}) and (\ref{equ12}) two pairs of
eigensolutions for the first derivatives of $v(x)$ and $\alpha(x)$
across the void boundary as a critical curve. The one that ensures
positive enclosed mass is
\begin{eqnarray}
&&v'\mid_{x^*}=\Big(-2+ 2\sqrt{2}\Big)(n-1)\ , \nonumber\\
&&\alpha'\mid_{x^*}=\sqrt{2}\bigg[\frac{(2-n)}{(3n-2)q}
+\frac{1}{2}\bigg](n-1)^2\frac{n}{hx^*}\ .\label{equ21}
\end{eqnarray}
It can be shown that for $n\neq 1$, all points on this critical
curve are saddle points (e.g. Jordan \& Smith 1977). It is known
that around a saddle singular point, only solutions along the
direction defined by eigensolutions are allowed. Therefore for
$\alpha^*=0$ at the void boundary, the self-similar solutions would
follow the behaviour described by expression (\ref{equ21}).
%Since $\alpha^*=0$, we should require a positive $\alpha'\mid _{x^*}$
%locally to ensure a physical solution; for this reason, we only take
%the plus sign in both expressions of (\ref{equ21}).{\bf Please clarify}
The presence of magnetic field is crucial here. In a purely
hydrodynamic case of $h=0$ or a conventional polytropic case of
$q=0$, the reduced density gradient $\alpha'$ diverges at the void
boundary $x=x^*$. It would require additional considerations for a
void boundary on which the mass density vanishes but its first
derivative diverges. To better describe a sudden change of mass
density on the void boundary, we would then set a non-zero mass
density $\alpha^*\neq 0$ at the void boundary (see analyses in the
following sections). The valid regime of parameters in which
solution (\ref{equ21}) stands is therefore $q>0,\ \gamma+q\geq 1$
and $h> 0$.
%\textbf{\textit{
As we should further require $\gamma\geq 1$ for a positive enthalpy,
the valid regime of parameters is simply $q>0$ and $h> 0$.

If the three functional coefficients ${\cal X}$, ${\cal A}$ and
${\cal V}$ do not vanish at the void boundary, the leading terms of
the first derivatives of $v$ and $\alpha$ at the void boundary are
then
\begin{eqnarray}
&&v'\mid _{x^*}=2(1-n)+(2-3n)q/\gamma\
,\label{CC1}\\
&&\alpha'\mid _{x^*}=\frac{(2-3n)q}{\gamma}\frac{\alpha}{(nx-v)}
\label{CC2}\nonumber \\
&&\quad\quad=-\frac{q\alpha}{(\gamma+q)}(x-x^*)^{-1}\ .
\end{eqnarray}
Note the coefficient $C$ disappears in these two expressions. The
second equality for $\alpha'$ in equation (\ref{CC2}) involves
equation (\ref{CC1}). The asymptotic solution of $\alpha$ near the
void boundary is
\begin{equation}
\alpha=K(x-x^*)^{-q/(\gamma+q)}+\cdot\cdot\cdot\ ,\label{CC3}
\end{equation}
where $K$ is an arbitrary integration constant referred to as the
void density parameter. To ensure solution (\ref{CC3}) going to zero
in the limit of $x\rightarrow x^*$,
%$x\rightarrow0^+$, {\bf You mean $x^*$?}
we require $q/(\gamma+q)<0$. We now verify that for such solutions,
${\cal X}$, ${\cal A}$ and ${\cal V}$ actually diverge
%do not vanish {\bf diverge?}
at the void boundary. With solutions (\ref{CC1})$-$(\ref{CC3}), we
have ${\cal X}\sim(x-x^*)^{q/(\gamma+q)}$, ${\cal
V}\sim(x-x^*)^{q/(\gamma+q)}$, ${\cal A}\sim(x-x^*)^{-1}$.
Therefore, the condition $q/(\gamma+q)<0$ in the meantime ensures
the validity of the asymptotic solution. We should require
$\gamma\geq 1$
%$\gamma>0$ {\bf [actually $\gamma\geq 1$]}
as a physical requirement and thus inequality $q/(\gamma+q)<0$ is
equivalent to $q<0$ and $\gamma+q>0$. %With solutions (\ref{CC1})
%and (\ref{CC2}), according to equation (\ref{equ19}), the first
%derivative of enclosed mass $dm/dx$ vanishes on the void boundary
%and the solution is smooth across the void boundary.

A positive enclosed mass requires $v'\mid _{x^*}<n$, which
%{\bf Please clarify.}
provides a lower limit for the index $q$, namely
\begin{equation}
q>q_{\rm min}=\frac{4+3n^2-8n}{3n(3n/2-1)}=\frac{2(n-2)}{3n}\
,\label{equ31}
\end{equation}
where $n-2<0$. %Numerical values of this $q_{\rm min}$ are tabulated
%in Table \ref{tab2}.
With algebraic manipulations, it is proven that
inequality (\ref{equ31}) together with $\gamma\geq 1$
%$\gamma>0$ {\bf [actually $\gamma\geq 1$]}
is equivalent to inequality $\gamma+q>0$. Thus within the parameter
regime of $q<0$ and $\gamma+q>0$ where solutions
$(\ref{CC1})-(\ref{CC3})$ stand, inequality $v'\mid _{x^*}<n$ is
automatically satisfied.
%
%\begin{table*}
% \centering
%  \caption{Values of $q_{\rm min}$ as defined by equation (\ref{equ31})
%  for void solutions to exist given various values of $n$}
%  \begin{tabular}{|c|c|c|c|c|c|c|c|}
%  %\hline
%  \hline
%   $n$ & 2/3 & 0.7 & 0.8 & 0.9 & 1.0 & 1.1 & 1.2  \\
%   \hline
%   $q_{\rm min}$ & $-4/3$
%   %$-\infty$
%   & $-1.238$ & $-1$ & $-0.815$ & $-2/3$ & $-0.545$ &
%   $-0.444$ \\
%      %\hline
%      \hline
%   $n$ & 1.3 & 1.4 & 1.5 & 1.6 & 1.7 & 1.8 & 1.9\\
%      \hline
%   $q_{\rm min}$& $-0.359$ & $-0.286$&$-0.222$&$-0.167$&$-0.118$&$-0.074$&$-0.035$\\
%      \hline
%\label{tab2}
%\end{tabular}
%\end{table*}

\begin{figure}
 \includegraphics[width=0.5\textwidth]{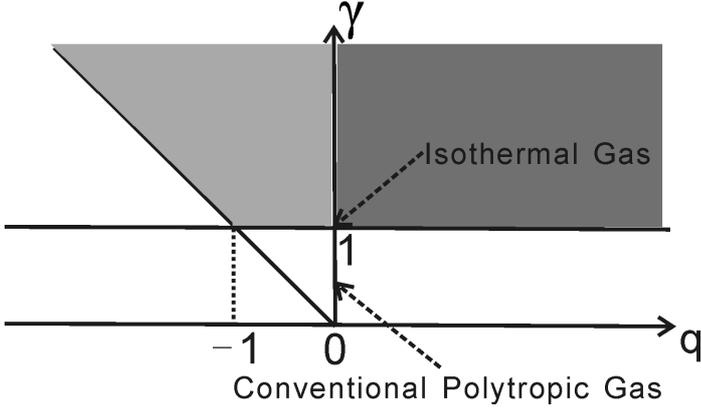}
 \caption{
 Regimes of parameters $q$ and $\gamma$ for which asymptotic solutions
 near the void boundary are applicable. The horizontal axis is the
 parameter $q$ and the vertical axis is the polytropic index $\gamma$.
 The solid line is for $\gamma+q=0$. The vertical axis of $q=0$
 corresponds to the conventional polytropic case ($n+\gamma=2$) and
 the point $q=0\ ,\ \gamma=1$ ($n=1$) corresponds to the isothermal
 case. The light shaded zone in the second quadrant is the regime to apply
 asymptotic solution $(\ref{CC1})-(\ref{CC3})$ and the heavy shaded zone
 in the first quadrant is the regime to apply asymptotic solution
 (\ref{equ21}). We only draw the $\gamma>0$ part out of physical
 consideration; in fact, we should always require $\gamma\geq 1$ to
 ensure a positive specific enthalpy. }
 \label{Regime}
\end{figure}

In summary, in the vicinity of the void boundary $nx-v=0$ with
$\alpha=0$, we find two possible types of asymptotic solutions. The
parameter regimes in which these solutions are applicable are shown
in Figure \ref{Regime} accordingly.
With $q>0$
%$\ \gamma+q-1\geqslant0$ {\bf [actually $\gamma\geq 1$]}
(the heavy-shaded zone in the first quadrant of Figure
\ref{Regime}) and $h>0$ (magnetized), the asymptotic solution
takes form (\ref{equ21}), referred to as LH1 solutions. With
$q<0,\ \gamma+q>0$ (the light-shaded zone in the second quadrant
of Figure \ref{Regime}), the asymptotic solution takes form
$(\ref{CC1})-(\ref{CC3})$, referred to as LH2 solutions. For other
regimes of parameters, no void solutions with boundary condition
$nx-v=0$ and $\alpha=0$ are found. The isothermal case and the
conventional polytropic case do not satisfy either of these two
requirements, so they need additional considerations for
asymptotic solutions near the void boundary with
$\alpha^*\neq0$ (e.g. Hu \& Lou 2008a; Lou \& Zhai 2009). %Lou \&
%Zhai (2009) have studied a hydrodynamic isothermal
%system and shown that void solutions satisfying $\alpha=0$ on the
%void boundary do not exist. Hu \& Lou (2008) study the void
%behaviour in the unmagnetized hydrodynamic conventional polytropic
%framework, and find that $\alpha^*=0$ can only give trivial solution
%as $\alpha=0$ for all $x$. These results are all compatible with the
%analysis here for the most general situations.
Hence, the novel LH1 and LH2 void solutions only exist in general
polytropic MHD cases with $q>0$ and $q<0$, respectively. LH1 void
solutions require the presence of magnetic field, while LH2
solutions remain valid in the purely hydrodynamic case as well.

\subsubsection[]{Hydrodynamics of
$\alpha^*> 0$ at the Void Boundary}

The situation with $\alpha^*\neq 0$ at the void boundary is
intrinsically different. In such cases, we require $q\geq 0$ to
ensure the pressure going to zero at the void boundary. One can
then numerically integrate coupled nonlinear ODEs (\ref{equ11})
and (\ref{equ12}) directly from the void boundary to construct
solutions. The possible singularity at the void boundary, and the
corresponding leading terms for ${\cal X}$, ${\cal A}$ and ${\cal
V}$ approaching the void boundary $nx-v=0$ depend largely on
parameter $q$. We find that $q=0$ for a conventional polytropic
gas is a special case giving a different expression of the
asymptotic behaviour at the void boundary. We examine below two
situations of $q=0$ and of $q>0$ separately. %We emphasize that for a
%general polytropic gas, the case of $n=1$ does not necessarily mean
%an isothermal gas as $\gamma>1$ is allowed.
%Hence later on we discuss the sub-situation with $q\geqslant 1$,
%$q<1,q\neq 0$ and $q=0$. We find that sometimes, the presence of
%a random magnetic field plays an important role.

%{\bf The Case of $q\geqslant 1$}

%If the reduced mass density $\alpha$ has a nonzero finite value on
%the void boundary, there is no singularity in the ODEs (\ref{equ11})
%and (\ref{equ12}), and a numerical integration from the void
%boundary can be carried out in a straightforward manner.

{\bf Case of $q=0$}

In the $q=0$ case, the formulation is simplified considerably and
we return to the conventional polytropic case of $n+\gamma=2$. For
non-magnetic cases of $h=0$, the asymptotic solution approaching
the void boundary is
\begin{eqnarray}
&&v=nx^*+2(1-n)(x-x^*)+\cdots\ ,\label{equ29c1}\\
\nonumber \\
&&\alpha=\alpha^*+\frac{n(1-n)}{\gamma}\alpha^{*n}x^*(x-x^*)+\cdots\
,\label{equ29c2}
\end{eqnarray}
where $x^*$ denotes the void boundary in a self-similar expansion
and $\alpha^*$ denotes the reduced mass density on the expanding
void boundary. This solution is the same as the void solution in
Hu \& Lou (2008a). Expression (\ref{equ29c2}) hints that, if
$\alpha^*=0$, then the solution becomes $\alpha(x)=0$ everywhere
for all $x$, and this is indeed so. Again, $\alpha^*=0$ is not
allowed for a conventional polytropic gas. In this case, no
apparent singularity is found near the void boundary, and both
terms $(nx-v)$ and $(\alpha-\alpha^*)$ scale as $(x-x^*)$, which
we denote as type-N (Normal) void behaviour. Note that LH1 void
solutions represent also a type-N void behaviour.

Series expansion solutions (\ref{equ29c1}) and (\ref{equ29c2})
become insufficient for $n=1$. In this isothermal case of $q=0$
and $n=1$ ($\gamma=1$), we can obtain asymptotic solutions near
the void boundary to a higher order. The leading terms of $v'$ as
$x\rightarrow x^*$ yield
\begin{equation}
v'\mid_{x^*}=\frac{2}{x^*}(x-v)\ .
\end{equation}
We then obtain the leading terms of $v$ and $\alpha$ as
$x\rightarrow x^*$
\begin{equation}
v=x^*+\frac{1}{x^*}(x-x^*)^2+\cdots ,\label{equ29e5}
\end{equation}
\begin{equation}
\alpha=\alpha^*-\frac{\alpha^{*2}}{2}(x-x^*)^2+\cdots .
\label{equ29e6}
\end{equation}
No singularity appears in the isothermal case (as the Type-N
behaviour) but the term $(nx-v)$ has the leading order of
magnitude $(x-x^*)^{2}$, which we refer to as the type-N2 void
boundary (see Lou \& Zhai 2009 for isothermal voids in
self-similar expansion).

%In cases of a magnetofluid with magnetic parameter $h\neq 0$, we
%readily have
%\begin{eqnarray}
%&&v=nx^*+2(1-n)(x-x^*)\ ,\label{equ29c3}\\
%&&\alpha=\alpha^*-\frac{\alpha^*[(n-1)n
%+2h\alpha^*]x^*}{(2-n)(\alpha^*)^{1-n}+h\alpha^* x^*}(x-x^*)\
%.\label{equ29c4}
%\end{eqnarray}
%By setting $h=0$, asymptotic solutions (\ref{equ29c3}) and
%(\ref{equ29c4}) reduce to equations (\ref{equ29c1}) and
%(\ref{equ29c2}).

{\bf Cases of $q>0$}

In such cases, the leading terms of functional coefficients ${\cal
X}$, ${\cal A}$ and ${\cal V}$ are also the thermal pressure terms
as in equation (\ref{XAV}), and then $v'\mid _{x^*}$ and
$\alpha'\mid _{x^*}$ %can be obtained from coupled nonlinear ODEs
%(\ref{equ11}) and (\ref{equ12}),
%\begin{eqnarray}
%&&v'\mid _{x^*}=\frac{(1-n)(4-2n+3nq)-nq}{2-n+(3n-2)q/2}\ ,
%\label{equ29}
%\\ \nonumber \\
%&&\alpha'\mid_{x^*}=-\frac{q\alpha} {(\gamma+q)}(x-x^*)^{-1}\
%.\label{equ29b}
%\end{eqnarray}
%Expressions (\ref{equ29}) and (\ref{equ29b})
are the same as expressions (\ref{CC1}) and (\ref{CC2}).
Therefore, the asymptotic form of $\alpha$ approaching the void
boundary is the same as equation (\ref{CC3}), viz.
$\alpha=K(x-x^*)^{-q/(\gamma+q)}$ where $K$ is the void density
parameter. The first derivative of $v$ in these cases tends to a
certain value on the void boundary and the term $nx-v$ has the
leading order of magnitude $x-x^*$.

As $q>0$ in this case, $\alpha$ diverges on the void boundary.
Here a sharp discontinuity in mass density exists across the void
boundary. We verify in turn that the divergence of $\alpha$ does
not affect the leading terms of functional coefficients ${\cal
X}$, ${\cal A}$ and ${\cal V}$ and the validity of expressions
(\ref{CC1}) and (\ref{CC2}). Despite this divergence, $m(x)$
remains continuous at $x=x^*$ (equation \ref{equ8}) and the
asymptotic solution with such singularity is physically allowed.
We refer to such asymptotic solution on the void boundary as the
type-D (diffusion) void behaviour.

We may regard the void boundary as a translation of centre along
the streamline $nx-v=0$. Previous asymptotic solutions at $x=0$
give either a zero $\alpha$ or a divergent $\alpha$ obeying power
law (e.g. $\alpha\propto x^{-3/2}$ for free-fall solutions, see
Lou \& Wang 2007). Here on a void boundary, the power law index of
the asymptotic $\alpha$ depends on parameter $q$. In terms of
physics, a local diffusion process may smooth out this
singularity, bearing in mind that self-similar behaviours will be
modified by local diffusions near the void boundary in
self-similar expansion. Relevant comments on this may be found in
Lou \& Cao (2008)
%\citet{b5}
and Lou \& Zhai (2009).

With $q=0$ in expressions (\ref{CC1}) and (\ref{CC2}), we have the
same $v'\mid _{x^*}$ as equation (\ref{equ29c1}) and $\alpha^*=K$.
In fact, the $q=0$ case (Type-N; $\alpha$ tends to a positive
constant) is a transitional case between $q<0$ (LH2, $\alpha$ tends
to zero at the void boundary) and $q>0$ (Type-D; $\alpha$ diverges
at the void boundary).
%\textbf{\textit{
So far we have provided sensible void solutions for purely
hydrodynamic cases with different $q$ values. For all such
solutions, the thermal pressure force becomes dominant near the void
boundary. Without magnetic field, the thermal pressure is the key
factor in determining the dynamics near the void boundary as it
should be.

%{\bf Stop reading here}

\subsubsection[]{MHD Cases with $\alpha^*\neq0$ at the Void Boundary}

In such cases, we first require $q\geq 0$
%$q\geqslant 0$
to ensure the pressure approaching zero at the void boundary. With
$h>0$ in equations (\ref{equ11}) and (\ref{equ12}), the leading
terms in the vicinity of a void boundary is different from the
hydrodynamic case. In the presence of magnetic field, the magnetic
force becomes dominant at the void boundary and diffusion behaviours
(Type-D) of the void boundary do not appear. The void boundary
generally shows no singularity in these MHD cases. There are four
distinct situations described below in different regimes of $q$
parameter.

{\bf Case $q=0$ for a conventional polytropic gas}

In such cases of conventional polytropic MHD, additional terms
associated with the magnetic force appear in the asymptotic
solution, having a similar form parallel to hydrodynamic expressions
(\ref{equ29c1}) and (\ref{equ29c2}),
\begin{eqnarray}
&&\!\!\!\!\!\!\!\!\! v=nx^*+\frac{2(\alpha^*)^{(1-n)}(1-n)\gamma}
{(\alpha^*)^{(1-n)}\gamma+h\alpha^*(x^*)^2}(x-x^*)+\cdots\ ,\label{equ29m1}\\
\nonumber \\
&&\!\!\!\!\!\!\!\!\!
\alpha=\alpha^*-\frac{n(n-1)\alpha^*+2h(\alpha^*)^2}
{\gamma(\alpha^*)^{(1-n)}+h\alpha^*(x^*)^2}x^*(x-x^*)+\cdots\
.\label{equ29m2}
\end{eqnarray}
Setting $h=0$ in solutions (\ref{equ29m1}) and (\ref{equ29m2}), we
retrieve solutions (\ref{equ29c1}) and (\ref{equ29c2}) as a
necessary check. This MHD solution manifests a type-N void behaviour
and the type-N2 void behaviour of the isothermal case disappears.

{\bf Cases of $0<q<1$}

In such cases, the presence of magnetic field becomes the leading
term approaching the void boundary, and the first derivatives of $v$
and $\alpha$ are respectively
\begin{equation}
v'\mid_{x^*}=\frac{2(1-n)}{\alpha^*}\ ,\label{equ29m3}
\end{equation}
\begin{equation}
\alpha'\mid_{x^*}=C\frac{q(2-3n)}{h}
(\alpha^*)^{1-n+3nq/2}(x^*)^{2q-2}(nx-v)^{q-1}\ .\label{equ29m4}
\end{equation}
Equation (\ref{equ29m4}) gives the leading term of the asymptotic
solution of $\alpha$ near the void boundary as
\begin{eqnarray}
\!\!\!\!\!\!\alpha=\alpha^*
+C\frac{(2-3n)}{h}(\alpha^*)^{\gamma}(x^*)^{2q-2} \qquad\qquad\\
\nonumber \qquad\qquad\times (\alpha^*n+2n-2)^{q-1}(x-x^*)^q\cdots
.\label{equ29m5}
\end{eqnarray}
To ensure the validity of this solution, we should require
$v'\mid_{x^*}<n$ for a positive mass.
%{\bf Why?}
For $n\geq 1$,
%$n\geqslant1$,
this condition is satisfied automatically, while for $n<1$, the
condition implies $\alpha^*n+2n-2>0$ or equivalently
$n>2/(2+\alpha^{\ast})$. No apparent singularity exists in this
solution and the term $(\alpha-\alpha^*)$ scales as $(x-x^*)^q$. We
refer to this asymptotic solution as type-Nq void behaviour.

{\bf Cases with $q=1$}

The asymptotic behaviour of $v$ near the void boundary is the same
as equation (\ref{equ29m3}), while the first derivative of $\alpha$
becomes
\begin{equation}
\alpha'\mid_{x^*}=\frac{(\alpha^*)^{1+n/2}
x^*(2-3n)-(n-1)n-2h\alpha^*}{hx^*}\ ,\label{equ29m6}
\end{equation}
giving a type-N behaviour near the void boundary. For $h\rightarrow
0$, the first derivative $\alpha'$ would diverge.

{\bf Cases with $q>1$}

The asymptotic behaviour of $v$ near the void boundary remains the
same as equation (\ref{equ29m3}), while the first derivative of
$\alpha$ becomes
\begin{equation}
\alpha'\mid_{x^*}=-\frac{(n-1)n+2h\alpha^*}{hx^*}\ ,\label{equ29m7}
\end{equation}
again giving a type-N behaviour near the void boundary.

%We have obtained various asymptotic solutions
%in the vicinity of the void boundary.
In different parameter regimes of $q$, $n$ (or $\gamma$) and $h$,
%with or without magnetic field,
we have obtained different types of asymptotic behaviours near the
void boundary and different expressions of asymptotic solutions. We
summarize these results in Table \ref{tab1} for reference.

\begin{table*}
%%\centering
\begin{center}
 %\begin{minipage}{140mm}
  \caption{Summary of asymptotic solution behaviours near the self-similar
  void boundary $nx-v=0$ and $\alpha\neq0$.
  In each case, we show the behaviour type and then equation numbers
  of the corresponding asymptotic solution. The properties of different
  types of solutions on the void boundary are described below the table.}
  \begin{tabular}[]{llll}
  \hline
   $q$ &$h=0$&$h\neq 0$\\
   \hline
   \hline
$q=0$& Type-N, (\ref{equ29c1}, \ref{equ29c2}); Type-N2
(\ref{equ29e5}, \ref{equ29e6})
& Type-N, (\ref{equ29m1}, \ref{equ29m2})\\
\hline $0<q<1$& Type-D, (\ref{CC1}, \ref{CC2})
& Type-Nq, (\ref{equ29m3}, \ref{equ29m4})\\
\hline $q=1$& Type-D, (\ref{CC1}, \ref{CC2})
& Type-N, (\ref{equ29m3}, \ref{equ29m6})\\
\hline $q>1$& Type-D, (\ref{CC1}, \ref{CC2})
& Type-N, (\ref{equ29m3}, \ref{equ29m7})\\
  \hline
\end{tabular}\label{tab1}
\begin{itemize}
\item Type-N: $\alpha$ tends to a nonzero finite value; and $(\alpha-\alpha^*)$ and $(v-v*)$ scale
as $(x-x^*)$.
\item Type-N2: $\alpha$ tends to a nonzero finite value; and $(\alpha-\alpha^*)$ and $(v-v*)$ scale
as $(x-x^*)^2$.
\item Type-Nq: $\alpha$ tends to a nonzero finite value; $(\alpha-\alpha^*)$ scales as
$(x-x^*)^q$; and $(v-v^*)$ scales $(x-x^*)$.
\item Type-D: $\alpha$ diverges and scales as $(x-x^*)^{-q/(\gamma+q)}$; and
$(v-v^*)$ scales $(x-x^*)$.
\end{itemize}
\end{center}
\end{table*}

\subsection[]{Asymptotic Self-Similar Solutions at Large $x$}

Prior studies have revealed various asymptotic self-similar
solutions of a quasi-spherical magnetofluid under self-gravity. At
small $x$, we have derived quasi-magnetostatic asymptotic solutions
(Lou \& Wang 2006, 2007; Wang \& Lou 2007 with $q=0$; Wang \& Lou
2008 with $q\neq 0$),
%(\citealt{b11,b12,b3} with $q=0$; \citealt{b4} with $q\neq 0$),
central MHD free-fall solutions (Shu 1977
%\citealt{b1}
for an isothermal gas;
%\citealt{b2}
Suto \& Silk 1988 for a conventional polytropic gas;
%\citealt{b4}
Wang \& Lou 2008 for a general polytropic gas), strong-field
asymptotic MHD solutions
%\citealt{b48}
(Yu \& Lou 2005 for an isothermal gas; Lou \& Wang 2007
%\citealt{b12}
for a conventional polytropic gas;
%\citealt{b4}
Wang \& Lou 2008 for a general polytropic gas). At large $x$, we
have asymptotic MHD solutions described below.
%in following subsections.

\subsubsection[]{Asymptotic MHD Solutions of Finite Density\\ \qquad\quad
 and Velocity in the Regime of Large $x$}

In this case, the gravitational force, the magnetic force (i.e. the
magnetic pressure and tension forces together), and the thermal
pressure force are in the same order of magnitude at large $x$. The
asymptotic solutions at large $x$ are given by Wang \& Lou (2008),
%\citet{b4},
namely
\begin{eqnarray}
\!\!\!\!\!\!\!\!&&\!\!\!\!\!\!\!\!\alpha=Ax^{-2/n}+\cdots\ ,\nonumber\\
\nonumber\\
\!\!\!\!\!\!\!\!&&\!\!\!\!\!\!\!\!v=Bx^{1-1/n}+
\bigg\{-\bigg[\frac{n}{(3n-2)}
+\frac{2h(n-1)}{n}\bigg]A\nonumber\\
&& \quad +2(2-n)n^{q-1}A^{1-n+3nq/2}\bigg\}x^{1-2/n}+\cdots\ ,
\label{equ23}
\end{eqnarray}
where $A$ and $B$ are two constants of integration, referred to as
the mass and velocity parameters respectively. To ensure the
validity of solution (\ref{equ23}), we require $2/3<n\leq 2$ (note
that inequality $n>2/3$ is directly related to self-similar
transformation (\ref{equ7}) and a positive enclosed mass). In case
of $2/3<n\leq 1$, the mass and velocity parameters $A$ and $B$ are
fairly arbitrary. In case of $1<n\leq 2$, velocity parameter $B$
should vanish to ensure that $v$ tends to zero at large $x$. This
valid range of scaling parameter $n$ corresponds to $\rho\propto
r^{-3}$ to $\rho\propto r^{-1}$. For the dynamic evolution of
protostellar cores in star-forming clouds, power-law mass density
profiles should fall within this range. This appears to be
consistent with observational inferences so far (e.g. Osorio, Lizano
\& D'Alessio 1999; Franco et al. 2000; McKee \& Tan 2002).
%\citep[e.g.][]{Osorio1999, Franco2000, Mckee2002}.
%{\bf Give a few references}

Furthermore by setting $v=0$ in MHD ODEs (\ref{equ11}) and
(\ref{equ12}), we readily obtain an exact global solution in a
magnetostatic equilibrium, namely
\begin{equation}
\alpha=A_0x^{-2/n}\ , \label{equ24}
\end{equation}
where the proportional coefficient $A_0$ is given by
\begin{equation}
A_0=\bigg[\frac{n^2-2(2-n)(3n-2)h}{2(2-n)(3n-2)}
n^{-q}\bigg]^{-1/(n-3nq/2)}\ .\label{equ24def}
\end{equation}
This describes a more general magnetostatic singular polytropic
sphere (SPS) with a substantial generalization of $q\neq 0$; the
case of $q=0$ or $n+\gamma=2$ is included here and corresponds to a
conventional polytropic gas of constant specific entropy everywhere
at all times (Lou \& Wang 2006, 2007; Wang \& Lou 2007, 2008).

\subsubsection[]{Asymptotic MHD Thermal Expansion Solutions}

At large $x$, the pressure may become dominant in certain situations
(Wang \& Lou 2008). Thus we may drop the magnetic and gravity force
terms in ODEs (\ref{equ11}) and (\ref{equ12}). By assuming $v\sim
cx+b$ and $\alpha \sim Ex^P$ with $c$, $E$ and $P$ being three
constant coefficients, ODEs (\ref{equ11}) and (\ref{equ12}) then
lead to
\begin{eqnarray}
&&P=-\frac{(3q-2)}{(1-n+3nq/2)}\ , \nonumber\\
\nonumber \\
&&E^{1-n+3nq/2}(n-c)^q(2+P)=c(1-c)\ , \nonumber\\
\nonumber \\
&&P=\frac{(3c-2)}{(n-c)}\ ,\label{equ25}
\end{eqnarray}
where the three constant coefficients $(c,\ E,\ P)$ can be
determined by equation (\ref{equ25}). This solution for MHD thermal
expansion is valid for $q>2/3$ as we need a power-law exponent $P<0$
for a converging $\alpha(x)$ at large $x$. We note that for a
certain system whose parameters are predefined, only one thermal
expansion solutions at large $x$ is allowed, except for a free
parameter $b$. Actually, a translation on $v$ will not alter the
structure of the solutions. %According to self-similar transformation
%(\ref{equ7}), the radial bulk flow speed can be expressed as
%\begin{equation}
%u=\frac{r}{t}\frac{v}{x}\ .\label{equ26}
%\end{equation}
The radial bulk flow speed at large $x$ is
\begin{equation}
u=c\frac{r}{t}\ . \label{equ27}
\end{equation}
At a certain time $t$, the radial flow speed is simply
proportional to $r$. There is a qualitatively similar flow speed
profile in the special case of $\gamma=4/3$ for a relativistically
hot gas (Goldreich \& Weber 1980; Lou \& Cao 2008; Cao \& Lou
2009).

\subsubsection[]{MHD Free-Expansion Solution}

In cases of $q<2/3$, numerical exploration suggests an expansion
solution in the asymptotic form of
\begin{equation}
v\rightarrow\frac{2}{3}x+b\ ,\qquad
\alpha\rightarrow\alpha_{\infty}\ ,\qquad\hbox{ as }
x\rightarrow+\infty\ ,\label{FreeE}
\end{equation}
where $\alpha_{\infty}$ is a constant value of $\alpha$ at large
$x$. With the radial velocity proportional to the radius in
asymptotic form (\ref{FreeE}) and $q<2/3$, the pressure gradient
terms in nonlinear MHD ODEs (\ref{equ11}) and (\ref{equ12}) can be
dropped and the leading terms of the three coefficients ${\cal X}$,
${\cal A}$ and ${\cal V}$ as $x\rightarrow+\infty$ are
\begin{eqnarray}
&&{\cal X}\sim h\alpha x^2-(nx-v)^2\ ,\nonumber\\
\nonumber \\
&&\!\!\!\!\!\!\!\!\!\!\!\!\!\!\!\!\!\!\!\!\!\! {\cal A}\sim
2\frac{x-v}{x}\alpha(nx-v)
%\nonumber\\
%&&\qquad
-\alpha\bigg[(n-1)v+\frac{(nx-v)}
{(3n-2)}\alpha+2h\alpha x\bigg]\ ,\nonumber\\
\nonumber \\
&&\!\!\!\!\!\!\!\!\!\!\!\!\!\!\!\!\!\!\!\!\!\! {\cal V}\sim
2\frac{(x-v)}{x}h\alpha x^2
%\nonumber\\
%&&\qquad
-(nx-v)\bigg[(n-1)v+\frac{(nx-v)}
{(3n-2)}\alpha+2h\alpha x\bigg]\ ,\nonumber\\
\label{XAV1}
\end{eqnarray}
respectively. To obtain asymptotic solution (\ref{FreeE}), we
require ${\cal A}\rightarrow0$ and ${\cal V}/{\cal
X}\rightarrow2/3$. The constant $\alpha_{\infty}$ then obeys the
following relation
\begin{equation}
(1+6h)\alpha^2-2\alpha/3=0\ .\label{alphacon1}
\end{equation}
Equation (\ref{alphacon1}) has only one non-trivial solution, namely
\begin{equation}
\alpha_{\infty}=\frac{2}{3(1+6h)}\ .\label{alphacon}
\end{equation}
We substitute this $\alpha_{\infty}$ into condition ${\cal V}/{\cal
X}\rightarrow2/3$ and find that this condition is satisfied. The
other solution $\alpha_{\infty}=0$ is indeed trivial and does not
satisfy condition ${\cal V}/{\cal X}\rightarrow2/3$. In summary, we
verify the existence of asymptotic expansion solution (\ref{FreeE})
in the regime $q<2/3$ and the constant value $\alpha_{\infty}$ is
given by equation (\ref{alphacon}). For such an expansion solution,
the pressure gradient is negligible, we thus refer to such expansion
solution as the `free-expansion' solution. The `free-expansion'
solution is the counterpart of thermal expansion solution for
$q<2/3$, as shown in this paper. The constant $\alpha_{\infty}$
depends not on parameter $n$, but on magnetic parameter $h$. With a
larger $h$ (i.e. a stronger magnetic field), the constant asymptotic
density is lower.

\subsubsection[]{The MHD Einstein-de Sitter Solution}

There exists a special exact semi-complete global solution referred
to as the MHD Einstein-de Sitter solution, having the form of
$v=2x/3$ and $\alpha=$ constant for all $x$. Wang \& Lou (2007)
%\citet{b3}
described this solution for the case of a conventional polytropic
magnetofluid (i.e. $q=0$). The form of this MHD Einstein-de Sitter
solution is
\begin{eqnarray}
&&v=\frac{2}{3}x\ , \qquad\qquad\qquad \alpha=\frac{2}{3(1+6h)}\ ,
\nonumber\\
\nonumber \\
&&m=\bigg(n-\frac{2}{3}\bigg)\frac{2x^3}{3(1+6h)}\ ,\qquad\qquad
q=0\ , \label{equ280}
\end{eqnarray}
where $n>2/3$. Compared with free-expansion solution ($\ref{FreeE}$)
and $\alpha_{\infty}$ value (\ref{alphacon}), we find that the
`free-expansion' solution naturally becomes the MHD Einstein-de
Sitter solution with $q=0$. We extend the consideration to the
general polytropic form adopted in this investigation. By setting
$v=2x/3,\ \alpha=$ const in nonlinear MHD ODEs (\ref{equ11}) and
(\ref{equ12}), it is clear that only $q=2/3$ (i.e. $\gamma=4/3$ with
an allowed range of $n$), in addition to the case of $q=0$, will
make the solution valid for all $x$.
%{\bf Sure of this? Impossible for other $q$?}
We have the novel MHD Einstein-de Sitter solution as
\begin{eqnarray}
&&v=\frac{2}{3}x\ ,\ \qquad
\alpha=\frac{2}{3}\frac{1}{6h+1+6C(n-2/3)^{2/3}}\ , \nonumber\\
&&m=\frac{2}{3}\frac{(n-2/3)x^3}{6h+1+6C(n-2/3)^{2/3}}\ , \qquad
q=\frac{2}{3}\ , \label{equ28}
\end{eqnarray}
where $n>2/3$.
%{\bf where $n>2/3$\ ? Please also provide a corresponding
%expression for the reduced enclosed mass $m$. Please
%also compare with the results of Lou \& Cao (2008)}.
Comparing with equation (80) of Wang \& Lou (2007),
%\citet{b3},
we have a more general form of Einstein-de Sitter solutions for a
relativistically hot gas, for which the gravity and magnetic forces
cannot be neglected with respect to the pressure force. Lou \& Cao
(2008) studied a similar relativistically hot gas of spherical
symmetry with another self-similar transformation and derive another
form of Einstein-de Sitter solution (36) of Lou \& Cao (2008). With
the freedom to choose $C$ parameter, which is linked with parameter
$C_0$ in Lou \& Cao (2008), these two forms of
Einstein-de Sitter solution are equivalent for $h=0$. %The
%dimensional relativistic version of this solution has been
%introduced as the background for the Hubble expansion of our
%Universe filled with collisionless dark matter particles in earlier
%studies on voids in cosmological contexts \citep[e.g.,][]{b34,b35}.
It is interesting to observe that such globally exact Einstein-de
Sitter solution can only exist either in a conventional polytropic
gas or in a relativistically hot gas.

We have explored above asymptotic expansion solutions in different
$q$ regimes and derived three kinds of expansion solutions.
%for the first time.
With $q<2/3$, the thermal pressure can be neglected and the free
expansion solution stands; with $q>2/3$, the gravitational and
magnetic forces can be neglected and the thermal-expansion solution
stands; with $q=2/3$, all the three forces are comparable and the
Einstein-de Sitter solution is an exact global solution. Here we see
that $q=2/3$ separates different situations of self-similar
expansion behaviour. All these expansion solutions have velocity
proportional to the radius. For the free-expansion and Einstein-de
Sitter solutions, $v=2x/3$ and $\alpha$ is equal to a certain
constant, while for the thermal-expansion solution, $v=cx$ and
$\alpha$ converges to zero at large $x$. We will see presently that
in general MHD void solutions merge into one kind of expansion
solutions (determined by $q$ parameter) far from the flow centre, if
the MCC is not encountered.

\section[]{Hydrodynamic and MHD Void Solutions with $\alpha^*=0$}

\subsection[]{Hydrodynamic Cases}

\begin{figure}
 \includegraphics[width=0.5\textwidth]{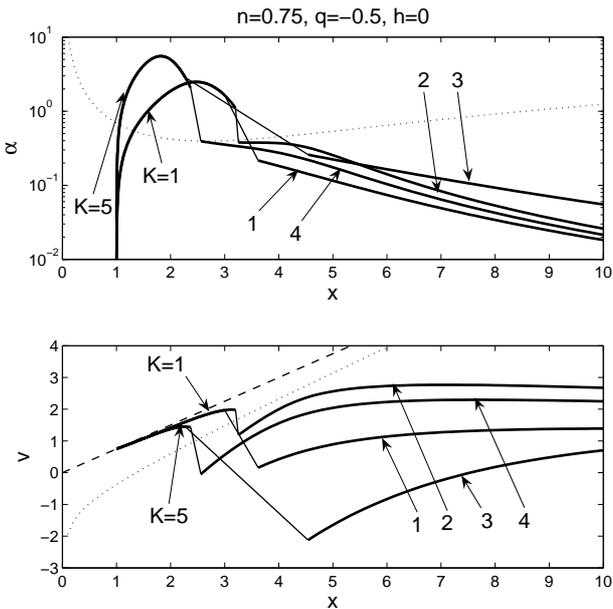}
 \caption{Global LH2 void solutions with shocks for parameters $(n=0.75,\
q=-0.5,\ \gamma=1.1875,\ h=0)$ and the void boundary $x^*=1$. The
upper panel shows the reduced density $\alpha(x)$ and the lower
panel shows the reduced radial velocity $v(x)$. In the upper panel,
a logarithmic scale is used for $\alpha(x)$. In both panels, the
dotted curve is the sonic critical curve, and in the lower panel the
dashed curve is the void boundary line $nx-v=0$.
%Relevant parameters are tabulated in Table
%\ref{tabpara} for a convenient reference.
The void solution with $K=1$ connects two solutions 1 and 2 whose
parameters are: $A=7.3025,\ B=3.7802,\ x_{\rm sd}=3,\ \alpha_{\rm
sd}=1.6366,\ v_{\rm sd}=1.9702,\ x_{\rm su}=3.6205,\ \alpha_{\rm
su}=0.2519,\ v_{\rm su}=0.1559$ (solution 1), and $A=8.7808,\
B=6.3723,\ x_{\rm sd}=3.2,\ \alpha_{\rm sd}=1.1040,\ v_{\rm
sd}=1.9755,\ x_{\rm su}=3.2529,\ \alpha_{\rm su}=0.3781,\ v_{\rm
su}=1.1854$ (solution2), respectively.
The void solution with $K=5$ connects two solutions 3 and 4 whose
parameters are: $A=24.493,\ B=4.854,\ x_{\rm sd}=2.27,\
\alpha_{\rm sd}=2.8499,\ v_{\rm sd}=1.4515,\ x_{\rm su}=4.5410,\
\alpha_{\rm su}=0.2587,\ v_{\rm su}=-2.1237$ (solution 3), and
$A=7.6878,\ B=5.4448,\ x_{\rm sd}=2.37,\ \alpha_{\rm sd}=2.0616,\
v_{\rm sd}=1.4308,\ x_{\rm su}=2.5603,\ \alpha_{\rm su}=0.3928,\
v_{\rm su}=-0.0456$ (solution4), respectively.
}
 \label{FigureK}
\end{figure}

In purely hydrodynamic cases with $h=0$, LH1 void solutions do not
exist and we consider only LH2 void solutions. The parameter regime
for LH2 void solutions is $q<0$ and $\gamma+q>0$. To illustrate LH2
void solutions by examples, we choose a set of parameters as
$(n=0.75,\ q=-0.5,\ \gamma=1.1875,\ h=0)$ and construct LH2 void
solutions with assigned values of void boundary $x^*$ and density
parameter $K$ of asymptotic form (\ref{CC1})$-$(\ref{CC3}). We
choose a downstream shock position $x_{\rm sd}$ and insert a
hydrodynamic shock there to match inner void solution with outer
asymptotic envelope solution (\ref{equ23}) of finite velocity and
density at large $x$. Note that with $n<1$ the velocity actually
tends to zero at large radii. Several such global LH2 void solutions
with shocks are shown in Figure \ref{FigureK}. We have also
performed numerical explorations with different parameter sets and
the results are qualitatively similar.

With $\alpha^*=0$ at the void boundary, the density first increases
and then decreases as $x$ increases; and the radial velocity
increases as $x$ increases. The density profiles of LH2 void
solutions (see the upper panel of Figure \ref{FigureK}) indicate a
prominent shell-type morphology surrounding a central cavity in
expansion. The peak density of the solution and the width of the
shell is modulated primarily by $K$ parameter, which varies for
different astrophysical gas flow systems. With different values of
$x_{\rm sd}$, different dynamic behaviours of the corresponding
upstream side can be obtained (e.g. the lower panel of Figure
\ref{FigureK}). In the vicinity of the upstream shock front, the
fluid can be either an inflow (solutions 1 and 2) or an outflow
(solutions 3 and 4). With adopted parameters, the fluid always
merges into an asymoptotic outflow (parameter $B>0$) far from the
system centre, whereas it possible to construct solutions with
$B\leq 0$ for LH2 voids (see MHD examples below and examples in
Figure \ref{Fig1} for a conventional polytropic gas).
% By properly choosing the shock position, LH2 void solutions can
%be connected with various dynamical behaviours of outer envelope.
Numerical explorations indicate that it is generally not possible to
make LH2 void solutions to cross the sonic critical curve smoothly.
In other words, the inclusion of hydrodynamic shocks is necessary in
order to construct sensible semi-complete global void solutions.

The possibility of asymptotic inflows at large $x$ associated with
central void solutions deserves special attention. Without shocks,
void solutions generally merge into asymptotic expansion solutions
with flow velocities remaining positive. This establishes the
physical links between central expanding voids and asymptotic
outflows. In the presence of shocks, the upstream may have
inflows, either near the shock front or sufficiently far from the
system centre. This means that initially and at large radii a
cloud system may involve inflow or contraction under the
self-gravity. When the central engine forms an expanding void, a
resulting shock may face the falling gas and expand outwards. For
example, in Hu \& Lou (2008a), the possibility of asymptotic
inflows at large $x$ is interpreted as a special scenario for
``champagne flows" in H II regions.

We emphasize that LH2 solutions here are the only void solutions in
non-magnetized cases with $\alpha^*=0$. The shell-type appearance is
a general feature for LH2 void solutions. Such solutions are
applicable to shell-type morphologies, widely observed in various
astrophysical gas systems, such as supernova remnants and hot
bubbles (e.g. Ferri\`erre 1998, 2001),
%\citep[e.g.][]{Fe2, Fe1},
H II regions (e.g. Hu \& Lou 2008a)
%\citep[e.g.][]{HL08},
and even cavities in galaxy clusters (e.g. McNamara et al. 2005).
%\citep[e.g.][]{McNamara}.
%{\bf Need a more precise description and relevant references.}
The increasing velocity with radius of such solutions suggests a
wind nature: the fast wind from the central cavity decelerates in
the shell and the mass is accumulated in the shell. This is
consistent with the picture of champagne flows of H II regions (e.g.
Hu \& Lou 2008a) and supernova remnants.

\subsection[]{MHD Cases}

\begin{figure}
 \includegraphics[width=0.5\textwidth]{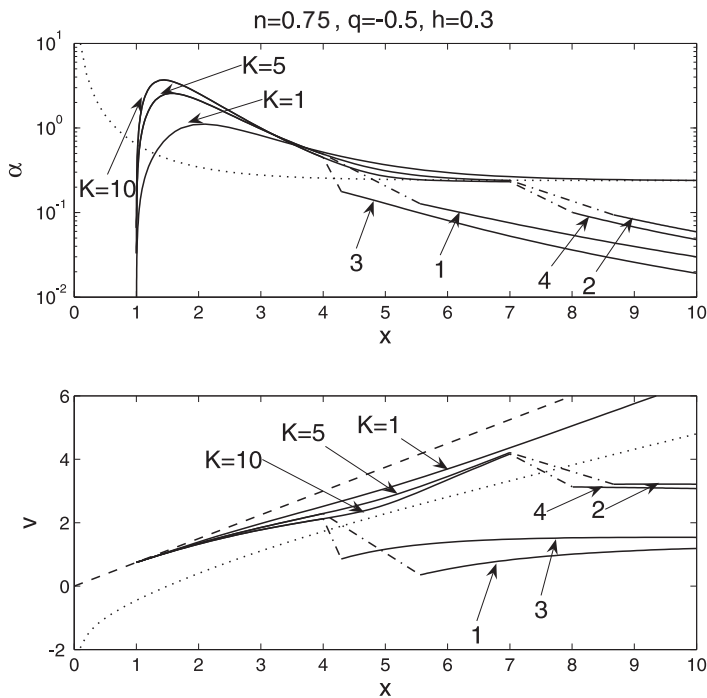}
 \caption{Global MHD LH2 void shock solutions with parameters $(n=0.75,\
q=-0.5,\ \gamma=1.1875,\ h=0.3)$ and a void boundary $x^*=1$. The
same format is adopted as in Figure \ref{FigureK}. The void
solution with $K=1$ does not allow a magnetosonic shock and it
merges into the MHD free-expansion solution. The void solution of
$K=5$ connects with solutions 1 and 2 whose parameters are:
$A=12.325,\ B=3.977,\ x_{\rm sd}=4,\ \alpha_{\rm sd}=0.4915,\
v_{\rm sd}=2.2888,\ x_{\rm su}=5.5577,\ \alpha_{\rm su}=0.1273,\
v_{\rm su}=0.3533$ (solution 1), and $A=18.390,\ B=8.344,\ x_{\rm
sd}=7,\ \alpha_{\rm sd}=0.2401,\ v_{\rm sd}=4.2102,\ x_{\rm
su}=8.6661,\ \alpha_{\rm su}=0.0941,\ v_{\rm su}=3.2135$ (solution
2), respectively.
The void solution of $K=10$ connects with solutions 3 and 4 whose
parameters are: $A=7.5520,\ B=4.010,\ x_{\rm sd}=4,\ \alpha_{\rm
sd}=0.4478,\ v_{\rm sd}=2.1273,\ x_{\rm su}=4.2913,\ \alpha_{\rm
su}=0.1775,\ v_{\rm su}=0.8558$ (solution 3), and $A=15.175,\
B=7.706,\ x_{\rm sd}=7,\ \alpha_{\rm sd}=0.2317,\ v_{\rm
sd}=4.1652,\ x_{\rm su}=8.0096,\ \alpha_{\rm su}=0.0999,\ v_{\rm
su}=3.1276$ (solution 4), respectively.
} \label{FigureKh}
\end{figure}

With a random magnetic field, both LH1 and LH2 void solutions exist.
As counterparts to hydrodynamic cases, we consider LH2 void
solutions with the same parameters adopted in the previous section,
except for the magnetic parameter being $h=0.3$. Such global MHD LH2
void solutions with shocks are shown in Figure \ref{FigureKh}.
%{\bf Something wrong with Figure 3 and Figure 4 for shocks!!}

The appearance of MHD LH2 void solution does not change very much
with a magnetic parameter $h>0$. The density profiles also show
shell-type morphology and the velocity still increases with radius
(see Fig \ref{FigureKh}). Compared with non-magnetized cases shown
in Figure \ref{FigureK} with the same void boundary $x^*$ and the
same parameter $K$, the peak density in the shell is lower and the
shell width appears broadened in MHD cases. The $K=1$ void solution
in the non-magnetized case can involve a shock, otherwise it
encounters the SCC. However, with the same value of $K$ in the
magnetized case, the void solution cannot harbor any shocks and
merge into the MHD free-expansion solution definitely. The $K=5$ and
$K=10$ void solutions in the magnetized case can harbor MHD shocks.
The MHD behaviour of the corresponding upstream sides, from our
numerical exploration, is all outflow (Solutions 1, 2, 3 and 4 of
Fig \ref{FigureKh}). With a larger $x_{\rm sd}$, or a faster shock,
the upstream outflow has larger velocity, both near the shock front
and far from the centre. The shell-type appearance is commonly
observed for LH2 void solutions, for hydrodynamic and MHD cases.

\begin{figure}
 \includegraphics[width=0.5\textwidth]{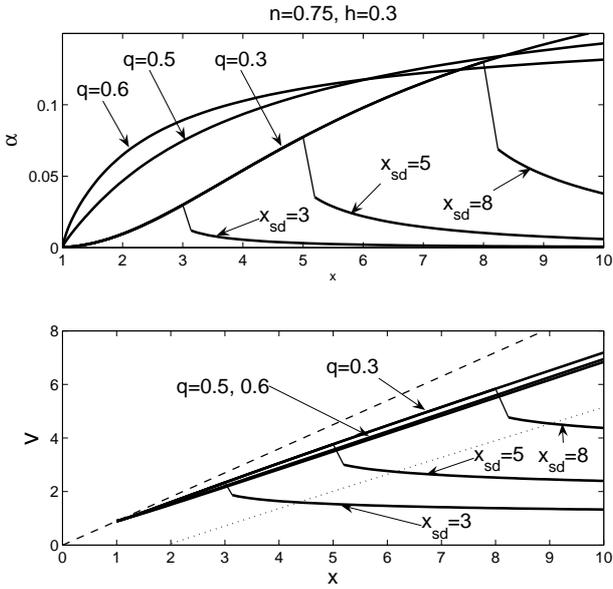}
 \caption{Semi-complete global MHD LH1 void solutions and shocks with the
parameter $(n=0.75,\ h=0.3)$ and the void boundary $x^*=1$. The
same format as Figure \ref{FigureK} is adopted. The void solutions
merge to the free-expansion solution (\ref{FreeE}) and the
constant value of $\alpha$ to be $\alpha_{\infty}=0.238$ (equation
\ref{alphacon}).
The void solution of $q=0.3$ connects with three upstream
solutions whose parameters are: $A=0.0968,\ B=1.6372,\ x_{\rm
sd}=3,\ \alpha_{\rm sd}=0.0299,\ v_{\rm sd}=2.3340,\ x_{\rm
su}=3.1374,\ \alpha_{\rm su}=0.0120,\ v_{\rm su}=1.8694$;
$A=0.9024,\ B=2.8925,\ x_{\rm sd}=5,\ \alpha_{\rm sd}=0.0774,\
v_{\rm sd}=3.7655,\ x_{\rm su}=5.1942,\ \alpha_{\rm su}=0.0353,\
v_{\rm su}=3.0033$; and $A=5.0384,\ B=5.2269,\ x_{\rm sd}=8,\
\alpha_{\rm sd}=0.1300,\ v_{\rm sd}=5.8366,\ x_{\rm su}=8.2401,\
\alpha_{\rm su}=0.0689,\ v_{\rm su}=4.7681$, respectively.
}
 \label{FigureLH1}
\end{figure}

We now consider MHD LH1 void solutions, for which the void boundary
$nx-v=0,\ \alpha=0$ is also a critical curve and the asymptotic
solution approaching the void boundary is an eigensolution. The
regime of parameter in which the LH1 void solution exists is $q>0,\
\gamma+q\geq 1,\ h>0$ ($\gamma\geq 1$). Examples of MHD LH1 void
solutions are shown in Fig \ref{FigureLH1}, which do not encounter
the critical curve and approach free-expansion asymptotic solution
(\ref{FreeE}) at large $x$.
%Unlike LH2 void solutions, LH1 void solutions do not encounter the
%magnetosonic critical curve and approach free-expansion asymptotic
%solution (\ref{FreeE}) at large $x$.
For this reason, the three velocity curves in the bottom panel of
Fig \ref{FigureLH1} are nearly identical. The density and velocity
increases as $x$ increases. With a larger parameter $q$, the density
profile appears more pronounced in the vicinity of void boundary.
Because the free-expansion solution has infinite velocity and
constant density far from the flow centre, the LH1 void solutions
would be more suitable for astrophysical model if they are matched
with another branch of solution with finite velocity and density at
large $x$ by MHD shocks. Examples of shocks are shown in Fig
\ref{FigureLH1} for $q=0.3$ with downstream shock positions $x_{\rm
sd}=3$, $x_{\rm sd}=5$, and $x_{\rm sd}=8$, respectively. Again with
a larger $x_{\rm sd}$, or a faster shock, the upstream outflow has a
higher speed.
%Such an example is shown in Figure \ref{FigureLH1} in the
%$q=0.5$ case with a downstream shock position $x_{sd}=9$.

\begin{figure}
 \includegraphics[width=0.5\textwidth]{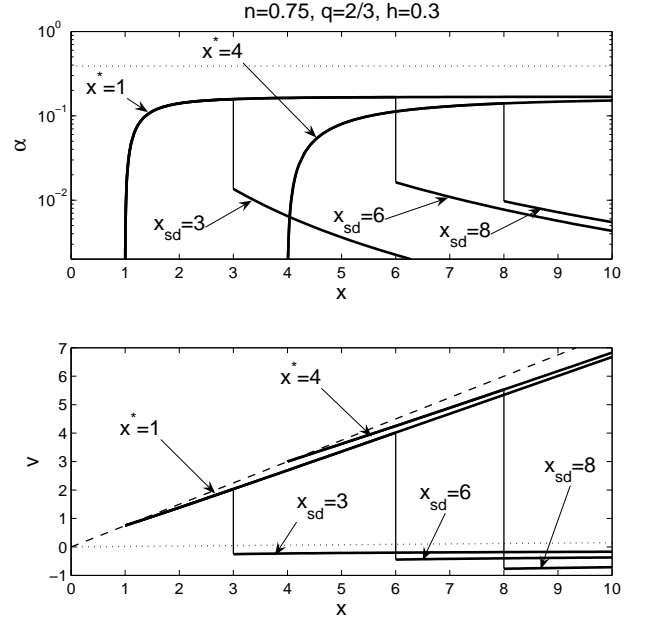}
 \caption{Semi-complete global MHD LH1 void shock solutions with
 parameters $n=0.75,\ q=2/3,\ \gamma=4/3,\ h=0.3,\ C=1$ and the
 void boundary $x^*=1$ and $x^*=4$. The same format as Figure
 \ref{FigureK} is adopted. The MCC is $\alpha=0.3897,\ v=0.0151x$.
The inner void solution of $x^*=1$ connects with two upstream
solutions whose parameters are: $A=0.2819,\ B=-0.3660,\ x_{\rm
sd}=3,\ \alpha_{\rm sd}=0.1577,\ v_{\rm sd}=2.0388,\ x_{\rm
su}=3,\ \alpha_{\rm su}=0.0136,\ v_{\rm su}=-0.2542$; and
$A=2.0920,\ B=-0.7974,\ x_{\rm sd}=6,\ \alpha_{\rm sd}=0.1666,\
v_{\rm sd}=4.0166,\ x_{\rm su}=6,\ \alpha_{\rm su}=0.0163,\ v_{\rm
su}=-0.4443$, respectively.
The inner void solution of $x^*=4$ connects to one upstream
solution whose parameters are: $A=2.7441,\ B=-1.5525,\ x_{\rm
sd}=8,\ \alpha_{\rm sd}=0.1404,\ v_{\rm sd}=5.5306,\ x_{\rm
su}=8,\ \alpha_{\rm su}=0.0097,\ v_{\rm su}=-0.7679$.
 }
 \label{FigureLH2}
\end{figure}

Examples of MHD LH1 void solutions for the relativistic case of
$q=2/3,\ \gamma=4/3$ are displayed in Fig. \ref{FigureLH2}. We set
free parameter $C=1$ and the arbitrary parameter $\lambda$ on the
shock to be $\lambda=1$.
%\textbf{
With equations (\ref{MCC231}) and (\ref{MCC232}), we obtain the MCC
with $\alpha=0.3897$ and $v=0.0151x$. In this case, the suitable
expansion solution becomes the Einstein-de Sitter solution as shown
in Fig. \ref{FigureLH2}. Similarly, we are able to construct MHD
shocks to match LH1 void solutions with another branch of solution
(\ref{equ23}) with finite velocity and density far from the flow
centre.

In general, the reduced density of MHD LH1 void solutions increases
with increasing $x$, and the density near the upstream side of shock
is very low, as compared with the density near the downstream side
of shock (see Figs. \ref{FigureLH1} and \ref{FigureLH2} for shocks).
For the corresponding upstream solutions, the density decreases as
$x$ increases and tends to zero at large $x$. With MHD shocks, we
obtain again the shell-type morphology for the density. This
shell-type morphology here is somehow different from the shell-type
morphology of LH2 void solutions (e.g. Figs. \ref{FigureK} and
\ref{FigureKh}). The LH2 void solutions have density peaks near the
void boundary by themselves: the density increases and then
decreases with increasing $x$. However, the LH1 void solutions must
involve MHD shocks to have shell-type density profiles and the peak
density is located on the downstream side of shock.

\section[]{Hydrodynamic Self-Similar Void Solutions with $\alpha^*\neq 0$}

By setting magnetic parameter $h=0$, we readily obtain a group of
self-similar nonlinear ODEs describing the hydrodynamics of a
general polytropic gas with specific entropy conserved along
streamlines. We shall choose a non-zero $\alpha^*$ at the void
boundary $nx-v=0$, or parameter $K$, in constructing solutions with
a considerable freedom. We also insert hydrodynamic shocks to obtain
semi-complete global solutions satisfying the asymptotic condition
that $v(x\rightarrow\infty)=0$. According to Table \ref{tab1},
asymptotic behaviours near the void boundary can be generally
classified as Type-N ($q=0,\ n\neq 1$), Type-N2 ($q=0,\ n=1$) and
Type-D ($q>0$) separately.

\subsection[]{Cases of $q=0$}

\begin{figure}
 \includegraphics[width=0.5\textwidth]{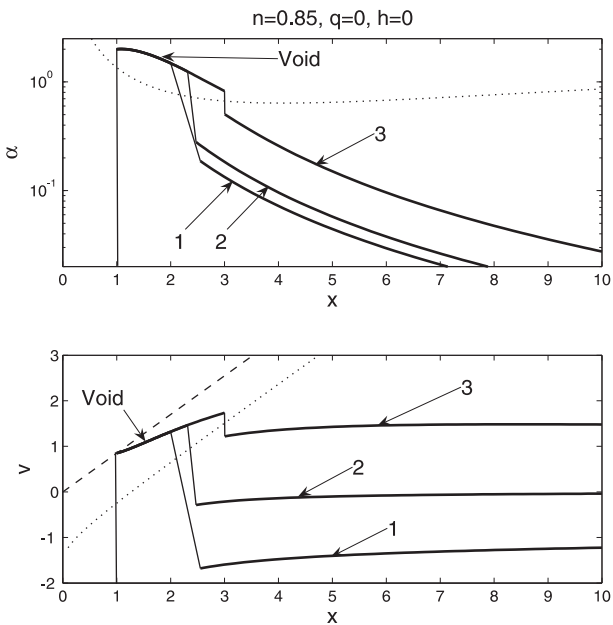}
 \caption{Void solutions with $n=0.85$, $\gamma=1.15$,
 $q=0$, $h=0$ for a conventional polytropic gas. The
 same format as Fig. \ref{FigureK} is adopted.
 The void solution of $x^*=1,\ \alpha^*=2$ connects with
 solutions 1, 2 and 3 whose parameters are: $A=2.2257,\ B=-1.8402,\ x_{\rm
sd}=2,\ \alpha_{\rm sd}=1.4849,\ v_{\rm sd}=1.3202,\ x_{\rm
su}=2.5521,\ \alpha_{\rm su}=0.1870,\ v_{\rm su}=-1.6783$
(solution 1); $A=2.5682,\ B=0,\ x_{\rm sd}=2.314,\ \alpha_{\rm
sd}=1.2452,\ v_{\rm sd}=1.4646,\ x_{\rm su}=2.4721,\ \alpha_{\rm
su}=0.2797,\ v_{\rm su}=-0.2876$ (solution 2); and $A=6.4646,\
B=3.2026,\ x_{\rm sd}=3,\ \alpha_{\rm sd}=0.8228,\ v_{\rm
sd}=1.7353,\ x_{\rm su}=3.0056,\ \alpha_{\rm su}=0.5033,\ v_{\rm
su}=1.2203$ (solution 3), respectively.
 }
 \label{Fig1}
\end{figure}

With $q=0$ and $h=0$, the flow system is reduced to a conventional
polytropic gas obeying $n+\gamma=2$ and have a constant specific
entropy everywhere at all times. Such hydrodynamic flows are
systematically and carefully analyzed and discussed by Wang \& Lou
(2007).
%\citet{b3}.
Hu \& Lou (2008a) constructed void solutions in a conventional
polytropic flow to model the so-called `champagne flows' in H II
regions. In Figure \ref{Fig1}, we present examples of void solutions
with hydrodynamic shocks across the sonic singular surface. For the
same void solution, by properly choosing the downstream shock
position $x_{\rm sd}$, we can obtain various dynamic behaviours on
the upstream side: outflow (e.g. $B>0$ in solution 3 of Figure
\ref{Fig1}), inflow (e.g. $B<0$ in solution 1 of Figure \ref{Fig1})
and contraction (e.g. $B=0$ in solution 2 of Figure \ref{Fig1}).
Basically, upstream dynamic behaviours depend on the void boundary
$x^*$, the density at the void boundary $\alpha^*$, and the
downstream shock position $x_{sd}$. Extensive numerical explorations
reveal that the isothermal case (Type-N2) is similar to other $q=0$
cases regarding the void solutions. Lou \& Zhai (2009) provide a
detailed analysis for isothermal voids.

\subsection[]{Cases of $q>0$}

\begin{figure}
 \includegraphics{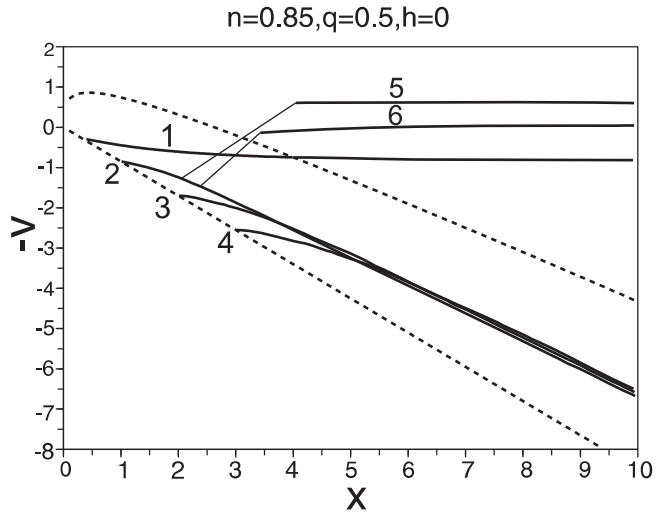}
 \caption{Inflow velocities $-v(x)$ of void solutions with $n=0.85$,
 $\gamma=1.2875$, $q=0.5$, $h=0$.
 %$-v$ is plotted.
 The upper dotted curve is the sonic critical curve (SCC), and the lower
dotted curve
 is the void boundary $nx-v=0$. The solid curve 1 is a void solution
 with $x^*=0.4$, crossing the upper SCC
 %sonic critical curve
 smoothly at $x=4$, and matching asymptotic solution (\ref{equ23})
 with $A=19.852$ and $B=1.286$. Solid curves 2, 3, 4 are integrated from
 $x^*=1$, $x^*=2$, $x^*=3$, with $K=0.29$, and $v'$ given
 by expression (\ref{CC1}). These solutions merge to the free expansion
 asymptotic solutions with form $v=2x/3+b,\ \alpha_{\infty}=2/3$.
 %Parameters of shocked solutions are tabulated in Table \ref{tabpara}.
%
 The void solution of $x^*=1$ and $K=0.29$ connects with solutions 5 and
 6 whose parameters are: $A=3.241,\ B=-0.987,\ x_{\rm sd}=2,\ \alpha_{\rm
sd}=0.502,\ v_{\rm sd}=1.246,\ x_{\rm su}=4.045,\ \alpha_{\rm
su}=0.114,\ v_{\rm su}=-0.607$ (solution 5), and $A=2.173,\
B=-0.143,\ x_{\rm sd}=2.4,\ \alpha_{\rm sd}=0.435,\ v_{\rm
sd}=1.479,\ x_{\rm su}=3.425,\ \alpha_{\rm su}=0.126,\ v_{\rm
su}=0.136$ (solution 6), respectively.
 }
 \label{Fig2}
\end{figure}

\begin{figure}
 \includegraphics{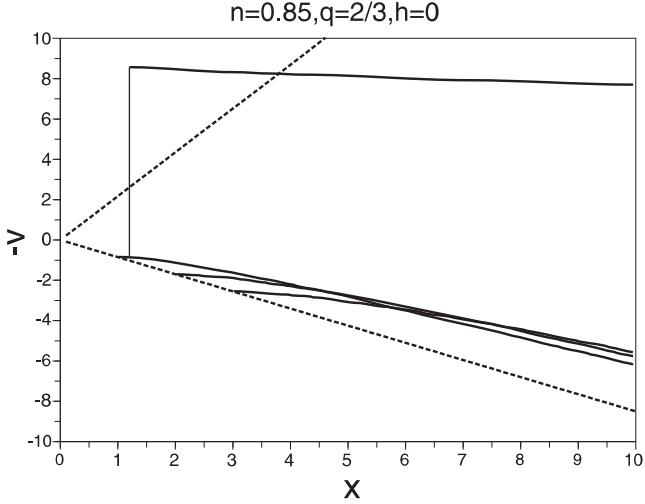}
 \caption{Void solutions and the Einstein-de Sitter solution with
 $n=0.85$, $\gamma=4/3$, $q=2/3$, $h=0$ and $C=1$. Inflow velocities
 $-v(x)$ are presented. The same format as Figure \ref{Fig2} is adopted.
 The critical curve is $v=-2.1772x$. The solid curves are integrated
 from the void boundary at $x^*=1$, $x^*=2$, $x^*=3$, with $K=0.2$ and
 $v'$ given by expression (\ref{CC1}), which match with the Einstein-de
 Sitter solution at large $x$.
The void solution with $x^*=1,\ K=0.2$ connects to one upstream
solution whose parameters are: $A=0.188,\ B=-12.131,\ x_{\rm
sd}=1.2,\ \alpha_{\rm sd}=3.453,\ v_{\rm sd}=0.874,\ x_{\rm
su}=1.2,\ \alpha_{\rm su}=0.0524,\ v_{\rm su}=-8.566$. }
 \label{Fig3}
\end{figure}

For $q>0$, asymptotic solution behaviours at the void boundary are
of Type-D. We can construct void solutions with shocks. Examples of
such void shock solutions with different values of $q$ are shown in
Figs \ref{Fig2} and \ref{Fig3}.

Dynamic behaviours of void solutions depend on the void boundary
$x^*$ and the density parameter $K$. Without encountering the sonic
critical point, some void solutions (see curves $2$, $3$ and $4$ in
Fig \ref{Fig2}) merge into one kind of asymptotic expansion
solutions (e.g. asymptotic free-expansion solution, with $v\sim
2x/3$ for $q<2/3$, Einstein-de Sitter solution for $q=2/3$, or
asymptotic thermal-expansion solution for $q>2/3$). As shown by
Figure \ref{Fig2}, void solutions with different void boundaries
merge into the same asymptotic free-expansion solution, except for a
slightly different $b$ parameter. Again, we shall connect these
solutions with another asymptotic solution of finite velocity and
density at large $x$ by shocks (see curves $5$ and $6$ in Fig.
\ref{Fig2}). By properly choosing the downstream shock position, we
could readily obtain an inflow ($v<0$) or an outflow ($v>0$) for the
upstream side of a shock. The radial flow velocity tends to zero at
large radii. Some void solutions cross the critical curve smoothly
(see curve 1 in Fig. \ref{Fig2}).

As shown in Fig. \ref{Fig3} of a relativistically hot gas, the
critical curve can be obtained analytically from equations
(\ref{MCC231}) and (\ref{MCC232}) with $\alpha=3.2843$ and
$v=-2.1772x$.
%}
%{\bf Please clarify and compare with Lou \& Cao (2008).}
%{\bf Should also mention Goldreich \& Weber (1980).}
Void solutions merge into the Einstein-de Sitter solution as
expected. The Einstein-de Sitter solution has a diverging velocity
at large $x$.
%, limiting its applications.
Again, this solution can be connected with an asymptotic solution
(\ref{equ23}) of finite velocity and density at large $x$ via
shocks.
%Here, we could freely set $\lambda$ parameter.

\section[]{MHD Self-Similar Void Solutions with $\alpha^*\neq 0$}

%In the presence of a random magnetic field with the magnetic
%parameter $h\neq 0$, MHD behaviours near the void boundary $nx-v=0$
%are quite different (see Table \ref{tab1}).
The magnetic force may play a key role in the vicinity of void
boundary and smooths out all divergence of the non-magnetized
cases. In a MHD flow ($q\geq 0$), $\alpha^*=0$ at the void
boundary is allowed for non-trivial void solutions with $h>0$,
e.g. MHD LH1 void solution. In this section, we compare
$\alpha^*=0$ and $\alpha^*>0$ cases, and see if the boundary value
of $\alpha^*$ modifies characteristic behaviours of void
solutions.

\subsection[]{Case of\ \ $q=0$ for a conventional polytropic gas}

\begin{figure}
 \includegraphics[width=0.5\textwidth]{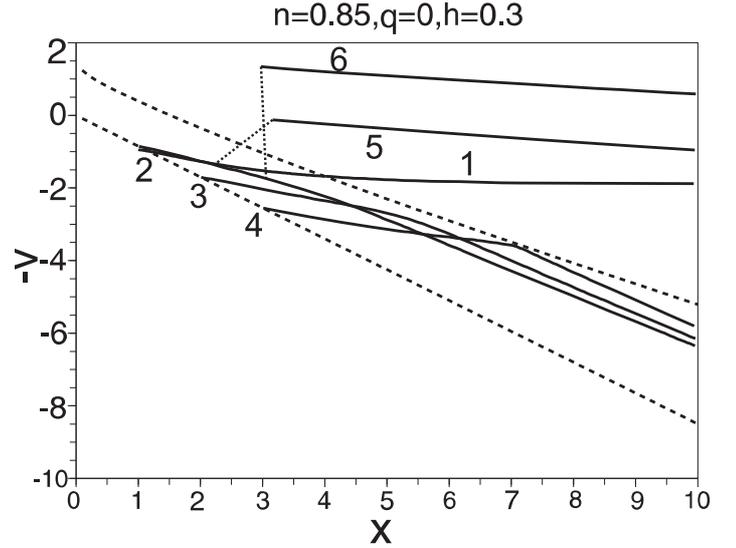}
 \caption{
 MHD void solutions with $n=0.85$, $\gamma=1.15$, $q=0$ and
 $h=0.3$. Inflow velocities $-v$ are plotted. The same format
 as Fig. \ref{Fig2} is adopted. Solid curve 1 is a void solution
 with $x^*=1$ and $\alpha^*=5$, crossing the MCC
 %magnetosonic critical curve
 smoothly at $x=4$ to match with asymptotic solution
 (\ref{equ23}) with $A=32$ and $B=5.413$ at large $x$.
 Solid curves 2, 3, 4 are integrated from the void boundary
 at $x^*=1$, $x^*=2$, $x^*=3$ respectively all with $\alpha^*=2$
 and $v'$ given by expression (\ref{equ29m1}). These solutions are
 limited by the upper MCC and merge into the Einstein-de Sitter
 solutions $(v=2x/3\ ,\ \alpha=\hbox{ const})$.
 The void solution of $x^*=1$ and $\alpha^*=2$ connects with
 solutions 5 and 6 whose
parameters are: $A=0.205,\ B=4.280,\ x_{\rm sd}=3,\ \alpha_{\rm
sd}=0.556,\ v_{\rm sd}=1.714,\ x_{\rm su}=3.159,\ \alpha_{\rm
su}=0.191,\ v_{\rm su}=0.123$ (solution 5), and $A=0.169,\
B=1.128,\ x_{\rm sd}=2.2,\ \alpha_{\rm sd}=0.868,\ v_{\rm
sd}=1.356,\ x_{\rm su}=3.306,\ \alpha_{\rm su}=0.126,\ v_{\rm
su}=-2.492$ (solution 6), respectively.}
 \label{Fig7}
\end{figure}

%With $q=0$ for a conventional polytropic magnetofluid, a MHD flow
%has a constant specific entropy everywhere and in time. In this
%special case, Einstein-de Sitter solution (\ref{equ280}) exists.
%According to expression (\ref{equ280}), the constant value of
%$\alpha$ depends on the magnetic parameter $h$.
Examples of global MHD void solutions for the case of $q=0$ are
shown in Fig. \ref{Fig7}. Numerical computations show that along the
MCC,
%magnetosonic critical curve,
velocity gradient $v'$ has one positive and another negative
eigenvalues, corresponding to the
%magnetosonic critical curve
MCC being saddle points (e.g. Jordan \& Smith 1977). Solutions
approaching this
%magnetosonic critical curve,
MCC, may either cross the critical curve smoothly and match with
asymptotic solution (\ref{equ23}) of finite velocity and density at
large $x$ (see solution curve 1 in Fig. \ref{Fig7}), or be turned
back smoothly to match with another branch of solutions and merge
into the Einstein-de Sitter solution at large $x$ (see solution
curves 2, 3 and 4 in Fig. \ref{Fig7}). By adjusting the $\alpha^*$
value at the void boundary, we can make the solution crossing the
%magnetosonic critical curve
MCC smoothly. The only difference to integrate curves 1 and 2 in
Fig. \ref{Fig7} is the $\alpha^*$ value. Compared with the case
without magnetic field (see Fig. \ref{Fig1}), a void solution in
this case can merge into the Einstein-de Sitter solution at large
$x$ without encountering the MCC.
%magnetosonic critical curve.
Void solutions can be connected with outer branch of the asymptotic
solutions of finite velocity and density by MHD shocks (see solution
curves 5 and 6 in Fig. \ref{Fig7}). Similarly by adjusting the
downstream shock position $x_{sd}$, one void solution on the
downstream side can be connected to various upstream solutions with
different behaviours at large $x$.

\subsection[]{Case of\ \ $q>0$}

\begin{figure}
 \includegraphics[width=0.5\textwidth]{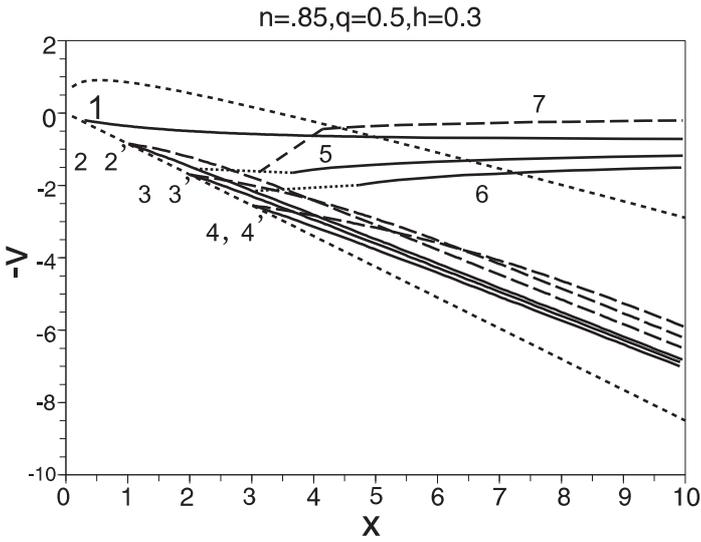}
 \caption{MHD void solutions with $n=0.85$, $\gamma=1.2875$, $q=0.5$,
 $h=0.3$. Inflow velocities $-v(x)$ are plotted. The same format as
 Fig. \ref{Fig2} is adopted. The solid curve 1 is a void solution
 with the void boundary at $x^*=0.3$, crossing the MCC
 %magnetosonic critical curve
 smoothly at $x=5$ to match with asymptotic solution
 (\ref{equ23}) with $A=25.977$ and $B=1.147$.
  The solid curves 2, 3, 4 are MHD LH1 void solutions,
 numerically integrated from the void
 boundary at $x^*=1$, $x^*=2$, $x^*=3$ with $\alpha^*=0$.
 Respectively, the dashed curves $2^{\prime}$,
 $3^{\prime}$, $4^{\prime}$ are Type-Nq void
 solutions, integrated from the void boundary
 at $x^*=1$, $x^*=2$, $x^*=3$ with $\alpha^*=2$.
 These solutions merge into the free expansion asymptotic solutions in the
 form of $v=2x/3+b$ at large $x$ with $\alpha_{\infty}=0.238$.
 The void solution 2 of $x^*=1, \alpha^*=0$ connects with
 solutions 5 and 6 whose
parameters are: $A=0.256,\ B=1.692,\ x_{\rm sd}=2,\ \alpha_{\rm
sd}=0.0646,\ v_{\rm sd}=1.475,\ x_{\rm su}=3.662,\ \alpha_{\rm
su}=0.0182,\ v_{\rm su}=1.648$ (solution 5), and $A=0.805,\
B=2.119,\ x_{\rm sd}=3,\ \alpha_{\rm sd}=0.0943,\ v_{\rm
sd}=2.142,\ x_{\rm su}=4.731,\ \alpha_{\rm su}=0.0299,\ v_{\rm
su}=1.991$ (solution 6), respectively.
The void solution 2' of $x^*=1$ and $\alpha^*=2$ connects with
solution 7 whose parameters are: $A=2.610,\ B=0.208,\ x_{\rm
sd}=3,\ \alpha_{\rm sd}=0.292,\ v_{\rm sd}=1.772,\ x_{\rm
su}=4.142,\ \alpha_{\rm su}=0.102,\ v_{\rm su}=0.445$. }
\label{Fig6}
\end{figure}

\begin{figure}
 \includegraphics[width=0.5\textwidth]{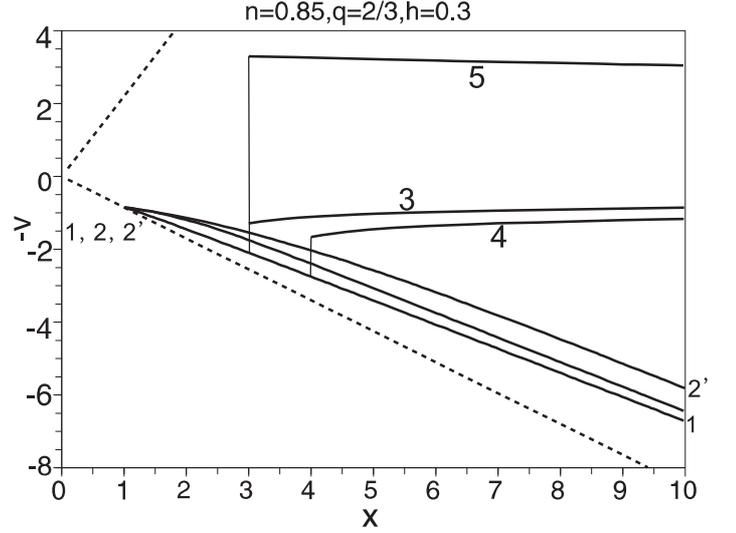}
 \caption{MHD void solutions with
 $n=0.85$, $\gamma=4/3$, $q=2/3$, $h=0.3$ and $C=1$. $-v$ is plotted.
 The upper dotted curve is the MCC satisfying $v=-2.2087x$ and the lower
 dotted curve is the void boundary $nx-v=0$. The solid curves 1 is
 LH1 void solution, integrated from the void boundary at $x^*=1$ with $\alpha=0$.
 The solid curves $2$ and $2'$ are Type-Nq void solutions, integrated
 from the void boundary at $x^*=1$ with $\alpha^*=2$ and $\alpha^*=10$.
 These solutions merge into the Einstein-de Sitter solution at large radii.
 The void solution 1 of $x^*=1, \alpha^*=0$ connects with solutions 3
 and 4 whose parameters are: $A=0.379,\ B=1.255,\ x_{\rm sd}=3,\ \alpha_{\rm
 sd}=0.114,\ v_{\rm sd}=2.095,\ x_{\rm su}=3,\ \alpha_{\rm
 su}=0.0413,\ v_{\rm su}=1.292$ (solution 3); and $A=0.848,\
 B=1.680,\ x_{\rm sd}=4,\ \alpha_{\rm sd}=0.123,\ v_{\rm
 sd}=2.750,\ x_{\rm su}=4,\ \alpha_{\rm su}=0.0459,\ v_{\rm
 su}=1.661$ (solution 4), respectively.
 The void solution 2' of $x^*=1$ and $\alpha^*=10$ connects with
 solutions 5 whose parameters are: $A=2.699,\ B=-4.921,\ x_{\rm
 sd}=3,\ \alpha_{\rm sd}=0.864,\ v_{\rm sd}=1.544,\ x_{\rm su}=3,\
 \alpha_{\rm su}=0.149,\ v_{\rm su}=-3.298$. }
 \label{Fig8}
\end{figure}

An example of $q=0.5$ is shown in Fig. \ref{Fig6}, which can be
compared with Fig. \ref{Fig2}. The case of $q=2/3$ describes a
relativistically hot gas. MHD void solutions in such case is shown
in Fig. \ref{Fig8} for a comparison with Fig. \ref{Fig3}.

Again the density at the void boundary $\alpha^*$ can be set to
either zero, which leads to the LH1 void solution with eigensolution
(\ref{equ21}) (see curves $2,\ 3$ and $4$ of Fig. \ref{Fig6}, and
curve 1 of Fig. \ref{Fig8}), or a nonzero finite value, which leads
to a Type-Nq behaviour of equations
(\ref{equ29m3})$-$(\ref{equ29m5}) (see curves $2^{\prime},\
3^{\prime}$ and $4^{\prime}$ of Fig. \ref{Fig6}, and curves $2$ and
$2^{\prime}$ of Fig. \ref{Fig8}). In the absence of a magnetic
force, only the Type-D behaviour is allowed in this range of $q$ and
the density diverges on the void boundary. If not encountering the
MCC, these solutions merge into one kind of expansion solutions at
large $x$ (i.e. free-expansion for $q<2/3$, Einstein-de Sitter
solution for $q=2/3$, and thermal-expansion for $q>2/3$). This
property is the same as hydrodynamic cases, and we note that the
constant density $\alpha_{\infty}$ of the free-expansion solution
does depend on the magnetic parameter $h$. This suggests that for
such void solutions, the magnetic force plays an important role.
Similar to non-magnetized cases, we can avoid the velocity
divergence of the expansion solutions by matching such solutions
with asymptotic solutions of finite velocity and density at large
$x$ via MHD shocks (e.g. curves $5,\ 6$ and $7$ of Fig. \ref{Fig6},
and curves 3, 4 and 5 of Fig. \ref{Fig8}), or making the void
solutions crossing the MCC smoothly (see curve 1 of Fig. \ref{Fig6}
and compare it with curve 1 of Fig. \ref{Fig2}). From Figs.
\ref{Fig6} and \ref{Fig8}, the LH1 void solutions and the Type-Nq
behaviour appear quite similar in terms of velocity profiles, except
near the void boundary and different $b$ parameter in the
corresponding expansion solutions. We will discuss the influence of
the initial mass density $\alpha^*$ presently.

The MCC has the form $\alpha=3.0088,\ v=-2.2087x$ for the case of
$C=1$, $\gamma=4/3$, $n=0.85$, $q=2/3$ and $h=0.3$. The behaviour of
MHD void solutions is similar to the hydrodynamic case. In such
cases, it is unlikely for void solutions to encounter the MCC and
the only way they cross this singular surface is via MHD shocks (see
curves $3,\ 4$ and $5$ in Fig. \ref{Fig8}).

We further investigate the influence of $\alpha^*$ value at the void
boundary on semi-complete global solutions. By comparing solutions
from the same void boundary $x^*$ and different $\alpha^*$ values
(see curves $1,\ 2,\ 2'$ in Fig. \ref{Fig8}), it appears that with
larger $\alpha^*$ on the void boundary, the void solutions converge
to the asymptotic solution more slowly. This means that with larger
density gradient on the void boundary, the system has a larger
transition zone where the magnetic force, the gravity and the
thermal pressure force are all comparable.
% in the same order of magnitude.
We refer to this zone as the void boundary layer. Outside the void
boundary layer, the free expansion, thermal expansion or the
Einstein-de Sitter solution would be a good approximation for the
asymptotic behaviour. With an MHD shock inserted, the solution can
be matched with an outflow (curves 3 and 4 in Fig. \ref{Fig8}) or an
inflow (curve 5 in Fig. \ref{Fig8}). The dynamic behaviour of the
outer upstream flow of the global void solution depends on the void
boundary $x^*$, the value of $\alpha^*$ and the downstream shock
point $x_{\rm sd}$. From the same void boundary $x^*$, we can adjust
$\alpha^*$ values to let the solution either cross the MCC smoothly
or merge into asymptotic expansion solutions (see curves 1 and 2 in
Fig. \ref{Fig7}).

\begin{figure}
 \includegraphics[width=0.5\textwidth]{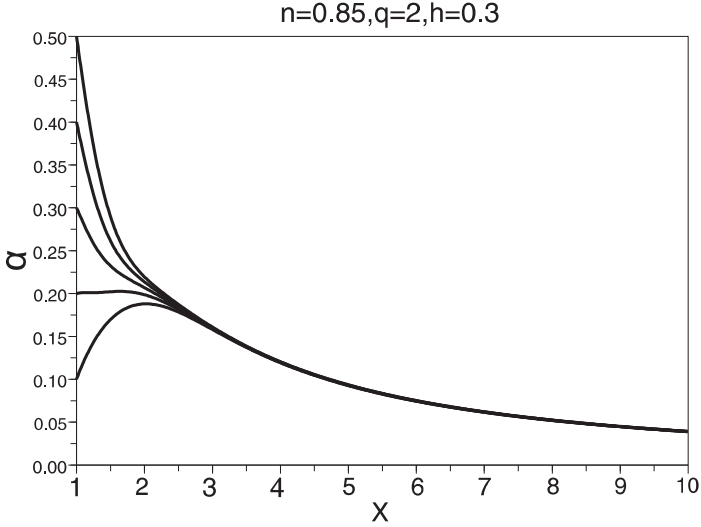}
 \caption{Influence of the initial value of $\alpha^*$ on the void
 solution. The five flow solution curves are integrated
 %in the system
 with $n=0.85$, $\gamma=1.7$, $q=2$, $h=0.3$, from the void boundary
 at $x^*=1$, and different initial values of
 $\alpha^*=0.5,\ 0.4,\ 0.3,\ 0.2,\ 0.1$ are chosen. The $v(x)$ part of these
 five solutions are nearly identical (not shown here). From this
 figure, for $x>2.5$, the five curves merge together, therefore the
 influence of the initial value of $\alpha^*$ is only significant around
 $1<x<2.5$.} \label{Fig5}
\end{figure}

As a further discussion, we obtain void solutions with different
$\alpha^*$ values at the void boundary with parameters $(n=0.85,\
\gamma=1.7,\ q=2,\ h=0.3)$ (see Figure \ref{Fig5}). In such cases,
there is no MCC.
%magnetosonic critical curve.
The solution curves are very similar and merge quickly as a single
curve (i.e. asymptotic thermal expansion solution). This suggests
that the general behaviour of the void solution is not influenced by
the initial $\alpha^*$ value. In other words, $\alpha^*$ at the void
boundary is fairly arbitrary and only influences the dynamical
behaviour near the void boundary (e.g. void boundary layer). For a
realistic astrophysical flow, the dynamics on the void boundary
cannot be described in a self-similar manner, due to unavoidable
diffusion processes. Therefore, choices of $\alpha^*$ at the void
boundary in our model serve only for starting a numerical
integration. A more complete understanding of such system requires
information for the relations between the $\alpha^*$ value on the
void boundary and the initial physical conditions that generate
voids, such as density perturbations and growths in supernova
explosions (e.g. Cao \& Lou 2009) or hot fast winds in planetary
nebulae (e.g. Lou \& Zhai 2009).

\section[]{Astrophysical Applications}

Our self-similar solutions can be adapted to different astrophysical
flow systems with various spatial and temporal scales. The sound
parameter $k$ determines the dimensional quantities in physical
space and $k$ varies for different astrophysical flow systems. From
self-similar transformation (\ref{equ7}), we have a relation
\begin{eqnarray}
k^{1-3q/2}=\frac{p}{(4\pi
G)^{\gamma-1}G^q(3n-2)^q\rho^{\gamma}M^q}\nonumber\\
=\frac{k_B T}{\mu (4\pi G)^{\gamma-1}G^q(3n-2)^q\rho^{\gamma-1}M^q}\
 \ ,\label{equk}
\end{eqnarray}
where $T$ is the thermal temperature, $k_B$ is Boltzmann's constant,
$\mu$ is the mean molecular (atomic) mass of gas particles and the
second equality only holds for an ideal gas. Since the entropy is
closely related to $p\rho^{-\gamma}$, the increase of entropy across
a shock front from the upstream side to the downstream side would
lead an increase of $k$ value in the same direction for $q<2/3$. For
$\gamma>1$ and $q<2/3$, the temperature should also increase across
a shock front in the same direction. It is still not trivial to
estimate $k$ values from relation (\ref{equk}) above, because the
enclosed mass $M$ varies in $r$. Numerical tests show that when $q$
is not too large, setting $q=0$ does not influence the magnitude
order of $k$ value. Thus we use a simplified relation
\begin{equation}
k=\frac{p}{\rho^{\gamma}(4\pi G)^{\gamma-1}}=\frac{k_B
T}{\mu\rho^{\gamma-1}(4\pi G)^{\gamma-1}}\ .\label{equk1}
\end{equation}
This is identical with relation (59) of Lou \& Wang (2006) for a
conventional polytropic gas. The relation does depend on the value
of $\gamma$. For a late evolution phase of massive stars after the
hydrogen burning, the central density and temperature are
$\rho_c\sim 10^8$ g cm$^{-3}$ and $T_c\sim 10^9$ K. We estimate
$k\sim10^{16}-10^{17}$ cgs unit, depending on the value of
$\gamma$. For the interstellar medium (ISM) in our own Galaxy,
mainly composed of hydrogen, $\rho_{\rm ISM}\sim10^{-20}-10^{-26}$
g cm$^{-3}$ and $T_{\rm ISM}\sim 10-10^6$ K (e.g. Ferri\`ere et
al. 2001) and we estimate $k\sim 10^{9}-10^{24}$ cgs unit,
depending on the value of $\gamma$ in the range of
$1\lsim\gamma\lsim 4/3$.

The parameters we have adopted in our model are $(n,\ \gamma,\ q,\
h)$ with the relation $\gamma=2-n+(3n-2)q/2$. The physical meaning
of these parameters is clear. Parameters $(n,\ \gamma,\ q)$ are
relevant for general polytropic processes.
%{\bf $q$ is proportional to the gradient of entropy perpendicular
%to streamlines.?} A non-zero $q$ indicates a fast heating or
%cooling process and the evolution is irreversible.
By setting $q=0$, we retrieve the conventional polytropic gas with a
constant specific entropy everywhere at all times and require
$n+\gamma=2$. The polytropic index $\gamma$ is an approximation
commonly invoked when energetic processes are not known (e.g. Weber
\& Davis 1967). For example, when applying our model to an exploding
stellar envelope, such as supernovae, $\gamma$ would be close to
unity, indicating a tremendous energy deposit. To apply our model to
a slowly-evolving ISM, $\gamma$ should be very close to ratio of
specific heats $c_{\rm p}/c_{\rm v}$ for an adiabatic process.
%the polytropic index in an adiabatic process.
This is consistent with the dynamic evolution shown by our
self-similar solutions. For a fixed $x$ value, the corresponding
radius $r$ expands with time obeying a power law of $\sim t^n$.
%Supposing $q=0$ and $\gamma<1$, {\bf We should
%require $\gamma\geq 1$!} we then have
For $n>1$, $r$ expands faster and faster, implying a continuous
energy input into a gas flow. Another role of $n$ is that it scales
the initial density distribution of a gas flow. According to
asymptotic solution (\ref{equ23}), the mass density scales as
$x^{-2/n}$ at large $x$. The initial condition with $t\rightarrow
0^+$ corresponds to the asymptotic boundary condition with
$x\rightarrow\infty$, so the scaling parameter $n$ determines the
initial density profile when the solution takes asymptotic form
(\ref{equ23}) at large $x$.

%\textbf{
Our general shock void solutions may be adapted to model planetary
nebulae. In the late stages of stellar evolution, the compact star
becomes an intense source of hot fast stellar wind and
photoionization. The fast wind catches up with the fully
photoionized slow wind and supports a fast wind bubble of hot gas.
%\citet{Chevalier1997}
Chevalier (1997) developed an isothermal self-similar model without
self-gravity to study the expansion of a photoionized stellar wind
around a planetary nebula (see also Meyer 1997). The key idea of
%\citet{Chevalier1997}
Chevalier (1997) is that the inner edge of the slow wind forms a
contact discontinuity with the stationary driving fast wind. We
have shown that this contact discontinuity in gravity-free cases
corresponds to the void boundary in our formulation. Lou \& Zhai
(2009) presented an isothermal model planetary nebula involving an
inner fast wind with a reverse shock; this shocked wind is
connected to an expanding self-similar void solution through an
outgoing contact discontinuity. In their model, the self-gravity
is included and a variety of flow profiles are possible. We here
provide a theoretical model formulation in a more general
framework with a polytropic equation of state and the inclusion of
self-gravity and the magnetic force. Our MHD shocked void
solutions are also suitable to describe the self-similar dynamics
of planetary nebulae combined with effects of central stellar
winds and photoionization.
%}

%\textbf{
Another astrophysical context to apply our MHD void shock solutions
is the expansion of H II regions surrounding new-born protostars,
especially for ``champagne flows" (e.g. Hu \& Lou 2008a).
%\citep[e.g.][]{HL08}.
Ultraviolet photons from nascent nuclear-burning protostars fully
ionize and heat the surrounding gas medium and drive H II regions
out of equilibrium. Such H II regions expand and gradually evolve
to a ``champagne flow'' phase with outgoing shocks.
%}

As a more detailed application of our void solutions, we revisit
below the scenario for core-collapse supernovae,
%and compare our model results with those of sophisticated
%numerical simulations by \citet{Janka1,Janka2}.
which has been investigated numerically over years (e.g. see
%\citet[][]{Lieben05}
Liebend\"orfer et al. 2005 for an overview). Neutrino-driven
models are widely adopted for explaining the physical mechanism of
type-II supernovae. The core collapse and bounce create a
tremendous neutrino flux and within several hundred milliseconds
after the core bounce, neutrino sphere is largely trapped and
deposit energy and momentum in the dense baryonic matter. A
typical scenario is that the neutrinos drive the stellar materials
outwards, deposit large amount of outward momentum and re-generate
the delayed rebound shock to push outwards.
%\citet{Janka1,Janka2}
Janka \& M\"uller (1995, 1996) successfully obtained numerical
simulation for the first second of type-II supernovae based on
such a scenario. Many recent numerical studies, with more careful
consideration of the convection in the stellar envelope and
diffusion processes, also confirm and consolidate the viability of
such a neutrino reheating process (e.g. Buras et al. 2006; Janka
et al. 2007, 2008).
%\citet{Buras06, Janka07, Janka08}.
From simulations of Janka \& M\"uller (1995, 1996)
%\citet{Janka1,Janka2}
on progenitor stars with a mass range of $\sim 8-15 M_{\odot}$,
the neutrino sphere stops depositing energy $\sim 0.5$ s after the
core bounce and then decouples from the baryonic matter. Once
decoupled, neutrinos quickly escape from the stellar interior and
may leave a cavity between the centre and the envelope within the
exploding progenitor star. At the centre of the cavity may lay a
nascent neutron star, a stellar mass black hole, or even shredded
debris (e.g. Cao \& Lou 2009).
%To make our void solutions applicable, we assume that the
%central compact remnant is a neutron star (this means the
%mass of progenitor should be less than $25M_{\odot}$).

We now show that the gravity of a remnant central object (if not
completely destroyed by the rebound process) on the expanding
stellar envelope may be neglected under certain situations. The
equivalent Bondi-Parker radius $r_{\rm BP}$ is
\begin{equation}
r_{\rm BP}=\frac{GM_*}{2a^2}\ ,\label{BP}
\end{equation}
%{\bf where is that factor 2?}
where $M_*$ is the mass of the central object and $a$ is the sound
speed of the surrounding medium, which mainly depends on
temperature. Far beyond $r_{\rm BP}$, the gravity of a central mass
may be ignored.
%\textbf{\textit{
In the following, we will see that at $\sim 1$ s after the core
bounce, the stellar envelope has a temperature of the order of $\sim
10^8$ K (see Fig. \ref{SNapp}) and a corresponding sound speed
squared $a^2\sim 10^{17}\hbox{ cm}^2\hbox{ s}^{-2}$. With $M_{*}\sim
M_{\odot}$, the Bondi-Parker radius at 1 s is estimated to be of the
order of $\sim 10^8$ cm which is roughly the same as the void
boundary. Therefore at the beginning, the gravity of the central
object is only marginally ignorable. From self-similar
transformation (\ref{equ7}), the Bondi-Parker radius, which is
proportional to $1/a^2\propto\rho/p$, has the time dependence of
$t^{2-2n}$. The void boundary expands as $t^n$. As long as we
require $n>2/3$, the Bondi-Parker radius expands slower than the
void boundary; in other words, the gravity of the central object may
be ignored shortly after the core bounce.
%}}
%For typical numbers, $M_{*}\sim M_{\odot}$ and
%$a^2\sim10^{19}\hbox{ cm}^2\hbox{ s}^{-2}$
%%{\bf unit correct?}
%(value obtained from the simulation), {\bf refs?
%implications?} we have $r_{\rm BP}\sim4\times10^6$ cm. The so-called
%`gain radius' in \citet{Janka1,Janka2}, the radius of the neutrino
%heated layer, which we believe indicates the scale of the central
%cavity at the moment of decoupling, is $\sim 100-125$ km at $t\sim
%50-100$ ms after the core bounce, already larger than $r_{\rm BP}$.
%With the envelope expansion, the cavity radius exceeds $\sim 10^8$
%cm within the first second of a supernova explosion.
We can then presume that the central cavity is an approximate void
and apply our self-similar void solutions. Similar approximation has
been applied to other astrophysical flow systems, in which the
gravity of central object can be neglected with respect to the
dynamics of the surrounding gas (e.g. see Tsai \& Hsu 1995; Shu et
al. 2002; Bian \& Lou 2005 and Hu \& Lou 2008 for applications to
shocked ``champagne flows" in H$_{\rm II}$ regions).

\begin{figure}
 \includegraphics[width=0.5\textwidth]{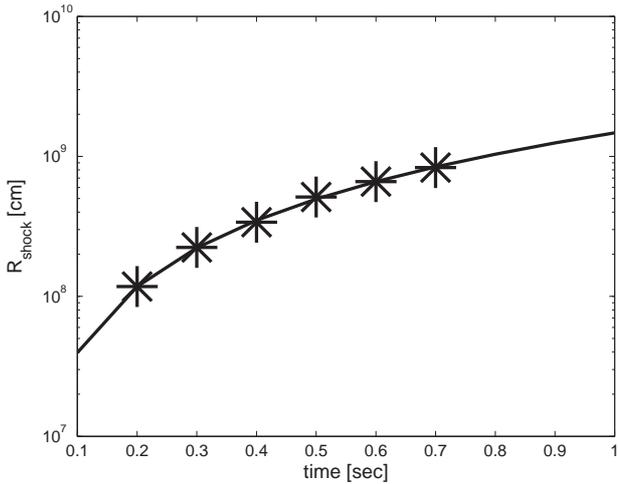}
 \caption{
 %{\bf We should have $\gamma\geq 1$.}
 Shock positions as function of time after a core bounce. The solid
 curve shows the self-similar evolution, with parameters
 $n=1.57,\ k=4\times 10^{16}$ cgs unit and the
 downstream shock position $x_s=7.36$. The asterisks show the result of
 numerical simulation by Janka \& M\"uller (1996)
 %\citet{Janka2}
 for a one-dimensional model of the core
 collapse of a progenitor with mass 15 M$_{\odot}$ and the iron core mass 1.31
 M$_{\odot}$, and the initial neutrino luminosity 2.225$\times 10^{52}$
 erg s$^{-1}$.}
 \label{Rshock}
\end{figure}

\begin{figure}
 \includegraphics[width=0.5\textwidth]{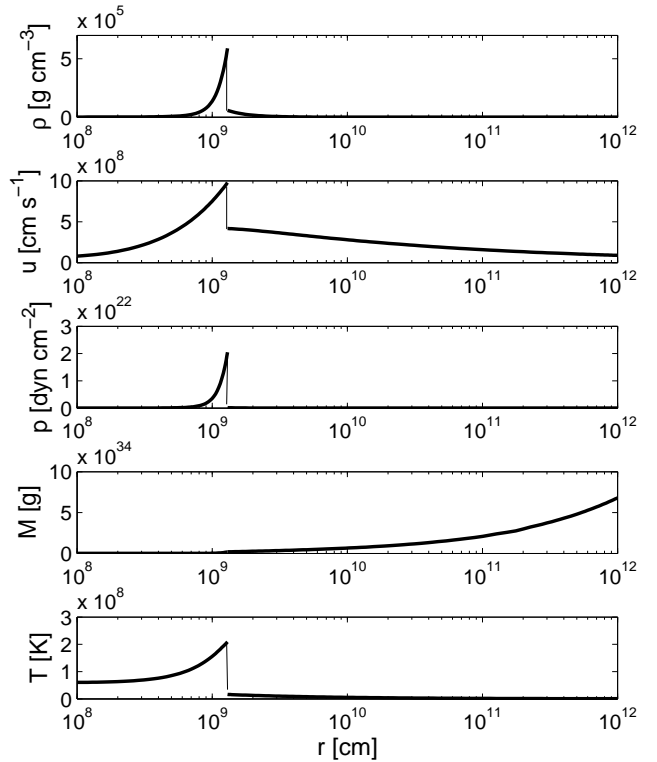}
 \caption{Self-similar void solution for an exploding progenitor star
 at $t=1$s. From top to bottom, the panels show the density, radial flow
 velocity, thermal pressure, enclosed mass and the thermal temperature.
 The solution is obtained with parameters $n=0.8,\ \gamma=1.2,\ q=0,
 \ h=0,\ k=4\times 10^{16}$ cgs unit, void boundary $x^*=0.5$, density
 at the void boundary $\alpha^*=0.001$ and the downstream shock position
 $x_{sd}=6.5$. Correspondingly at the moment shown, the void boundary is
 at $10^8$ cm and the rebound shock is at $1.3\times10^9$ cm. Our
 solution is numerically integrated until $10^{12}$ cm.}
 \label{SNapp}
\end{figure}

%{\bf Please carry out this comparison using a general polytropic
%model!} Within the first second after the core bounce, tremendous
%energy is deposited in the system, therefore we expect a small
%$\gamma$. {\bf We requre $\gamma\geq 1$!} To illustrate a simple
%application, we do not consider the magnetic field and constraint
%the parameters as $h=0,\ q=0$ and therefore $n+\gamma=2$.
%\textbf{\textit{
The model framework of self-similar dynamics described in this
paper implies that the radius of a spherical shock $r_s$ evolves
with time $t$ in a power-law manner, i.e. $r_s=k^{1/2}x_st^n$. We
may regard $k^{1/2}x_s$ and $n$ as two free parameters and attempt
to fit the self-similar evolution to the results of a numerical
simulation by Janka \& M\"uller (1996; e.g. their case O3c with
relevant parameters specified in the caption of our Figure
\ref{Rshock}).
%\citet[][e.g. their case O3c with relevant
%parameters specified in the caption of our
%Figure \ref{Rshock}]{Janka2}.
%}}
We use $k=4\times 10^{16}$ cgs unit for the inner part of the void
solution (i.e. the downstream side of an outgoing shock).
%We then compare the evolution of the rebound shock with time $t$,
%calculated from our self-similar model and that provided by the
%numerical simulation \citet[][e.g. Case O3c, with parameters
%specified in the caption of Figure \ref{Rshock}]{Janka2}.
The best fit model is achieved at $n=1.57$ and the downstream
shock position (or speed) of $x_s=7.36$ (see Fig. \ref{Rshock}).
%{\bf There will be an entropy problem to construct a shock.}
The self-similar evolution fits almost perfectly with the
simulation. It is striking to obtain such a good agreement for shock
evolution, because the numerical simulation of
%\citet{Janka2}
Janka \& M\"uller (1996) employed an equation of state that contains
contributions from neutrinos, free nucleons, $\alpha$-particles and
a representative heavy nucleus in nuclear statistical equilibrium.
In other words, their model carries distinct features.
%, and the equation of state adopted in \citet{Janka2}
%is fully vectorized. {\bf Meaning?}
With our simple approximation, we essentially parameterized all
these complicated energetic processes by a general polytropic
equation of state. We expect to obtain different best-fit scaling
parameter $n$, in comparison with numerical results under
different conditions, such as higher or lower initial neutrino
luminosity. This fitting is suggestive that the polytropic
approach is a fairly good approximation for shock evolution, and
physically the rebound shock expands in a self-similar manner.
%{\bf You need to invoke a more general
%polytropic process with $\gamma>1$.}

%\textbf{\textit{
The numerical simulation of Janka \& M\"uller (1996)
%\citet{Janka2}
ends at $\sim$ 1 s. Within this duration, neutrinos deposit enough
momentum and kinetic energy in a shocked stellar envelope and the
star is set to explode as the rebound shock emerges from the
stellar photosphere. The subsequent dynamic evolution, including
the travel of the rebound shock, can be readily described by our
self-similar model.
%}}
%{\bf Need to spell out a consistent story!}
Soon after neutrinos decouple from the baryon matter, no more
energy is provided and indeed the system begins to lose energy by
radiation processes. Thus, we can no longer apply $n=1.57$
further. We therefore use their model parameters at $\sim 1$ s as
the initial input parameters of subsequent dynamic evolution. We
found that if $n=0.8,\ \gamma=1.2,\ q=0$ (i.e. a conventional
polytropic gas), our self-similar model gives appropriate
solutions and we do not include a random magnetic field in this
preliminary illustration. The void solutions with these parameter
are of Type-N and relevant examples are also shown in Fig.
\ref{Fig1}. The cavity radius is taken to be $\sim 1000$ km at
$t=1$ s and the corresponding void boundary is then $x^*=0.5$.
From the simulation case O3c of Janka \& M\"uller (1996),
%\citet{Janka2},
the rebound shock is at $\sim 1.3\times 10^9$ cm at $t=1$ s, which
correspond to a downstream shock position $x_{sd}=6.5$. We note
that for these parameters the mass density cannot be set to zero
at the void boundary, hence we should choose $\alpha^*$ properly
such that the mass density on the void boundary at $t=1$ s is
equal to the mass density given by the case 03c. The void solution
at $t=1$ s is shown in Fig. \ref{SNapp}.

The solution shown in Figure \ref{SNapp} corresponds well to the
envelope of an exploding massive star. The enclosed mass
%{\bf rises as?}
is $\sim 25 M_{\odot}$ with a radius of $\sim 10^{12}$ cm, grossly
consistent with typical masses and radii of O and B stars.
%\textbf{\textit{
The enclosed mass mainly depends on the density $\alpha^*$ on the
void boundary; by regarding $\alpha^*$ as a free parameter, the
self-similar dynamics is capable of modelling stars with different
masses.
%}}
The solution shows an expansion velocity $\sim 10^9$ cm s$^{-1}$, a
typical expansion velocity for type-II supernovae. The temperature
increases from the void boundary to the rebound shock and decreases
with $r$ outside the rebound shock. The temperature rises to $\sim
10^8$ K, grossly consistent with typical supernova temperature.
%The density seems a bit small, as the density
%inside a massive supernova is $\sim 10^7-10^8$ g cm$^{-3}$.
%Actually, since the solution shows agreeable enclosed mass, the
%inconsistency on the density is minor. {\bf Please clarify}
Furthermore, we can compute the total energy of the entire system.
We should consider the kinetic energy $E_k$, gravitational energy
$E_g$ and the internal energy $E_i$. When calculating the internal
energy, we assume that the gas is single-atomic, whose degree of
freedom is 3. At $t=1$ s,
%{\bf unit?}
the total energy $E_{total}$ is $6.54\times 10^{50}$ erg, in which
the kinetic energy $E_k=1.63\times 10^{51}$ erg, the gravitational
energy $E_g=-1.10\times 10^{51}$ erg, and the internal energy
$E_i=1.21\times 10^{50}$ erg. We see that the kinetic energy is the
major energy source, and the radial motion of the fluid has not been
dissipated much to random motions, in agreement with the exploding
star scenario.
%{\bf Please provide specific estimates}
The total energy of our self-similar model is well consistent with
the total energy $9.5\times 10^{50}$ erg given by the Case O3c of
simulation (Janka \& M\"uller 1996)
%\citet{Janka2},
which suggests that the energy of the stellar envelope of the
self-similar model comes from the neutrino process before 1 s. Our
calculations reveal that the energy does not vary much with time,
consistent with our assumption that the system is nearly adiabatic.
We conclude that in general, our self-similar void solution is
plausible as it shows typical value of velocity, pressure,
temperature, enclosed mass and total energy.
%Noticeably, our analytical self-similar solutions, although
%approximate the energy process by a polytropic equation of state
%and do not consider the multi-dimensional convection in the
%stellar envelope, give fairly good agreement with the complete
%numerical simulation.
We can also adjust the self-similar parameters to produce solutions
for different objects.
%We will discuss the self-similar evolution of type-II supernovae in
%more details in a separate paper (Hu \& Lou 2008, in preparation).

Here we discuss two immediate utilizations of our self-similar
solutions. First, as long as we know the evolution of a rebound
shock, we can estimate the time when the shock breaks out of the
stellar envelope. Assuming the stellar photosphere at a radius of
$\sim 10^{12}$ cm, with the relation $r_{\rm shock}\propto t^n$, we
estimate that the shock travels to the photosphere at $\sim 4\times
10^3$ s (i.e. $\gsim 1$ hour) after the core rebounce. From then on,
we should be able to detect the massive star in act of an explosion.
%{\bf Recent Nature reference as an example}
With advanced instrument, astronomers are now detecting more and
more shock breakout events, and consolidate the association of long
$\gamma$-ray bursts (GRBs) and supernovae (e.g. Campana et al.
2006).
%\citep[e.g.][]{Campana}.
%\textbf{\textit{
From Figure \ref{SNapp}, the temperature around a shock in the
stellar envelope is in the range of $\sim 10^8$ K and gradually
decreases to the order of $\sim 10^7$ K as the shock breaks out of
the stellar atmosphere. Such a temperature range will give rise to
X-ray radiations. With our dynamical rebound shock model, coupled
with radiation (e.g. the thermal bremsstrahlung) and transfer
processes, we can calculate early X-ray emissions from supernovae in
act of an explosion (Lou \& Zhai 2009), and in turn, we may infer
properties of GRB/SN progenitor by observations of shock breakout
diagnostics. Hu \& Lou (2008b)
%\citet{HL08P}
presented some preliminary model calculations along this line and
found sensible agreement with X-ray observations of SN 2008D (e.g.
Soderberg et al. 2008; Mazzali et al. 2008).
%\citep{Soderberg2008, Mazzali}. %}}

Secondly, after long temporal lapses, we intend to relate the
central cavity of the stellar envelope with hot bubbles observed in
our own Galaxy. A typical SNR grows for $\sim 1.5$ Myr and reaches a
radius of $\sim 50$ pc (e.g. Ferri\`ere 2001). We assume that the
cavity and the stellar envelope (gradually evolving into a SNR)
expand in a self-similar manner as the solution shown in Figure
\ref{SNapp} after a SN breakout event. The cavity radius expands as
$t^n$, and 1.5 Myr later, the radius of the cavity becomes $\sim 3$
pc.
%(removed by Hu on April 24 according to referee)
%Despite the extreme simplicity of our estimate, we find
%that the result is not that far from the typical value.
%
Possible explanations for the discrepancy are: first, in the ISM the
sound scaling factor $k$ is much larger than that in the stellar
envelope; therefore we can no longer assume $k$ to be an overall
constant. Secondly, the typical scale of a SNR includes both the
cavity radius and the radius of the matter shell in the surrounding;
and the gas shell can well spread out to the ISM since the pressure
in the ISM is very low.
%Still, our self-similar solution provides
%a fairly good order-of-magnitude estimate.

As the last part of our discussion, we emphasize the shell-type
solutions such as LH1 void solutions and LH2 void solutions, and
corresponding examples are shown in Figs. \ref{FigureK},
\ref{FigureKh}, \ref{FigureLH1} and \ref{FigureLH2}. These void
solutions have zero density at the void boundary and thus a sensible
continuity across the void boundary. It is essential to consider the
magnetic field for the LH1 void solutions, as our study shows that
the magnetic field is indispensable to obtain shell-type LH1
solutions. As the shell-type SNRs are commonly observed, and the
magnetic field generally exists in SNRs, our self-similar void
solutions are likely to be a sensible dynamic approximation. In
cases with $q>2/3$, LH1 shell-type solutions merge into the
asymptotic thermal-expansion solution at large $x$ with a divergent
velocity. These solutions do not encounter the MCC,
%magnetosonic critical curve,
indicating that the fluid keeps sub-magnetosonic for the entire flow
system and such shell-type solutions cannot be matched with another
asymptotic solutions of finite velocity and density at large $x$ by
an MHD shock. In reality, the outer layer of shell-type SNRs is
bounded by the ISM.

\begin{figure}
 \includegraphics[width=0.5\textwidth]{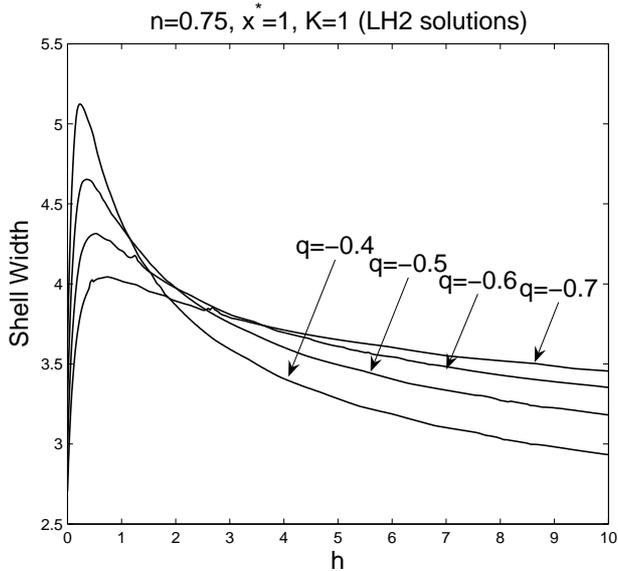}
 \caption{Shell width variation of LH2 solutions versus the magnetic
 parameter $h$ for different values of $q$ parameter. The shell width
 is defined as the distance from the void boundary to the place where
 the density is $e^{-1}$ of the peak density. The common parameters
 for these solutions are $n=0.75,\ x^*=1,\ K=1$.}
 \label{FigWidth}
\end{figure}

We define a shell width as the distance from the void boundary to
the place where the density is $e^{-1}$ of the peak density and
perform numerical exploration to examine how the shell width
depends on magnetic field strength $h$ and the self-similar
parameter $q$. One example of LH2 solutions with $n=0.75,\ x^*=1,\
K=1$ is shown in Figure \ref{FigWidth}. For certain $q$, the shell
width has a maximum value with $h\lsim 1$. The shell width
increases with $h$ rapidly in the regime $h\ll 1$ and decreases
with $h$ gradually in the regime of $h>1$. For a larger $q$ value,
the maximum shell width is larger, and the shell width at large
$h$ is smaller. In general, we find that the magnetic field and
the entropy distribution (or parameter $q$) influence the
shell-type morphology significantly. This implies that shell
structures in supernova remnants may reveal some information on
the magnetism of interstellar media or the internal energy input
or output of the gas. 

\section[]{Summary and Conclusions}

In a general polytropic MHD formulation, we obtain novel
self-similar void solutions for a magnetofluid under self-gravity
and with quasi-spherical symmetry. For our MHD void solutions, the
enclosed mass within the void boundary is zero and our solutions are
valid from the void boundary outwards. We have carefully examined
MHD and hydrodynamic behaviours near the void boundary and obtained
complete asymptotic solutions in various parameter regimes. For
clarity, we consider the situation of density being zero at the void
boundary and other situations separately.

For the case of $\alpha^*=0$ at the void boundary, we derive novel
LH1 solutions with $q>0$, for which the void boundary is also a
critical curve, and LH2 solutions with $q<0$, for which the thermal
pressure force becomes dominant approaching the void boundary. A
random magnetic field must be present in order to construct LH1
solutions. Both LH1 and LH2 solutions have shell-type morphology in
the density profile, whose peak density and the shell width are
mainly determined by the values of the magnetic parameter $h$ and
the parameter $q$. The shell width of the density profile expands
with time also in a self-similar manner.

For the case of $\alpha^*\neq 0$ at the void boundary, the situation
is quite different between the hydrodynamic and MHD cases. In
hydrodynamic cases, asymptotic behaviours in the vicinity of void
boundary depends on parameter $q$ and can be classified as Type-N
($q=0$) and Type-D ($q>0$). For the Type-D behaviour, the density at
the void boundary diverges, and local diffusion process should
occur. We systematically examined all these possibilities and
present a few numerical solution examples. Our solutions are well
compatible with previous self-similar solutions; for example, by
setting $q=0$, our formulation reduces to self-similar solutions for
a conventional polytropic gas (Lou \& Wang 2007).
%We find that different behaviours on the void boundary
%result from the competition among several forces.
Without magnetic field, for both
Type-N and Type-D behaviours, the thermal pressure force dominates
at the void boundary; actually Type-N solutions can be regarded as a
natural extension of LH2 solutions in the regime of $q=0$, and
Type-D solutions as a natural extension of LH2 solutions in the
regime of $q>0$. Therefore, for all sensible void solutions in
hydrodynamic framework, the thermal pressure is the dominant force
at the void boundary.
%For the Type-Q behaviour, the gravity force and the thermal
%pressure force are in the same order of magnitude on the void
%boundary. For the Type-Q2 behaviour, the gravity force dominates
%on the void boundary. {\bf Why is this?}
By including a random magnetic field, which is ubiquitous in
astrophysical plasmas, the divergence at the void boundary appearing
in the type-D behaviour can be removed, and the dominant force on
the void boundary becomes the magnetic force.

Void solutions may go across the critical curve either smoothly or
by an MHD shock and then merge into asymptotic solution
(\ref{equ23}) of finite density and velocity at large $x$. If void
solutions do not encounter the critical curve, they generally merge
into one kind of asymptotic expansion solutions at large $x$ with
the velocity proportional to the radius. For $q<2/3$, void solutions
merge into asymptotic free-expansion solutions, for which the
thermal pressure force is negligible. For $q>2/3$, void solutions
merge into asymptotic thermal-expansion solutions, for which the
thermal pressure force is dominant. For $q=2/3$, void solutions
merge into the Einstein-de Sitter solution, a semi-complete global
exact solution.
%In principle, void solutions can be matched with another
%branch of solutions and merge into asymptotic solutions
%of finite density and velocity by MHD shocks.
We are free to choose the position of void boundary $x^*$, the
density on the void boundary $\alpha^*$ (or the density parameter
$K$) and the downstream shock position $x_{\rm sd}$ to construct
various void solutions with different asymptotic dynamic behaviours
far from the void centre, including inflows, outflows, contraction
and breeze for upstream solutions.

%For $q>1$, the magnetosonic critical curve may have some special
%properties. We have shown examples of a critical curve in `portion'
%as well as `multiple' critical curves. In other words, for certain
%void solutions, it is impossible to establish MHD shocks and they
%are destined to merge into asymptotic thermal expansion solutions.

In this paper, we briefly discussed the case of $q=2/3$ and thus
$\gamma=4/3$ for a relativistically hot gas (Goldreich \& Weber
1980; Lou \& Cao 2008; Cao \& Lou 2009). One more parameter appears
in the self-similar form of the equation of state denoted as $C$ in
this case. We show solution examples of $C=1$. A more detailed
study, including cases of $C\neq 1$ is forthcoming.

Finally, we provide examples of applications of our self-similar MHD
void solutions. In principle, our solutions can be adapted to
various astrophysical plasmas with a central cavity. The scale of a
system can vary within the upper limit of neglecting the universe
expansion. The input and output of energy can be approximated by
properly choosing parameters $(n,\ \gamma,\ k)$. As more
self-consistent solutions of an MHD problem usually require
tremendous computational effort, our self-similar approach is
valuable in conceptual modelling and in checking simulation results.
We provide an application of our void solutions to the neutrino
reheating mechanism for core-collapse supernovae and compare the
dynamical results of self-similar solutions with previous numerical
simulations. We find that our simplified model fits well with
numerical simulations, suggesting that the self-similar approach is
plausible and the exploding stellar envelope and the rebound shock
do evolve in a self-similar manner. More specifically, we estimate
that a rebound shock breaks out from the stellar photosphere $\sim
4000$ s after the core bounce, for a progenitor star of mass 25
M$_{\odot}$ and radius $\sim 10^{12}$ cm. We expect that, our
general polytropic self-similar MHD solutions, coupled with
radiative transfer processes, may offer physical insight for the
rebound shock evolution of massive stars as well as on GRB-supernova
associations. We also indicate potential applications of LH1 and LH2
void solutions, with shell-type morphology, on the shell-type
supernova remnants as well as hot bubbles in the interstellar
medium.

\section{Acknowledgments}
This research was supported in part
%by the ASCI center for Astrophysical Thermonuclear
%Flashes at the University of Chicago,
by the Tsinghua Centre for Astrophysics,
%by the Collaborative Research Fund from the National Science
%Foundation of China (NSFC) for Young Outstanding Overseas
%Chinese Scholars (NSFC 10028306) at the National Astronomical
%Observatories, Chinese Academy of Sciences,
by the NSFC grants 10373009 and 10533020 and by the National Basic
Science Talent Training Foundation
%Guo Jia Ji Chu Ke Xue Ren Cai Pei Yang Ji Jin
%(often cited as Li Ke Ji Jin)
(NSFC J0630317) at Tsinghua University, and by the SRFDP 20050003088
and 200800030071 and the Yangtze Endowment from the Ministry of
Education at Tsinghua University. The kind hospitality of Institut
f\"ur Theoretische Physik und Astrophysik der
Christian-Albrechts-Universit\"at Kiel is gratefully acknowledged.

%{\bf Stop reading here}

\appendix
\section[]{MHD shock jump conditions}
\label{MHDshock}

MHD shocks can be constructed for self-similar solutions to cross
the MSS.
%magnetosonic singular surface.
Despite discontinuities in pressure, mass density, temperature,
magnetic field and velocity across the shock front, we require
conservations of mass, radial momentum, MHD energy across a shock
front in the comoving shock reference framework, respectively. They
are
\begin{equation}
\big[\rho(u_s-u)\big]_1^2=0\ ,\label{equ13}
\end{equation}
\begin{equation}
\bigg[p+\rho(u_s-u)^2+\frac{<B_t^2>}{8\pi}\bigg]_1^2=0\ ,
\label{equ14}
\end{equation}
\begin{equation}
\bigg[\frac{\rho(u_s-u)^3}{2}+\frac{\gamma p
(u_s-u)}{(\gamma-1)}+\frac{<B_t^2>}{4\pi}(u_s-u)\bigg]_1^2=0\ ,
\label{equ15}
\end{equation}
where $u_s$ is the shock travel speed in the laboratory framework
of reference. Since we only consider the dominant transverse
magnetic field parallel to the shock front in our theoretical
model framework, the magnetic induction equation can be written as
\begin{equation}
\big[(u_s-u)^2<B_t^2>\big]_1^2=0\ .\label{equ16}
\end{equation}
The magnetic field average is made over a layer between $r$ and
$r+dr$, and such average can still describe the discontinuity in the
radial direction. Strictly speaking, the magnetic fields have weak
radial components normal to the shock front. We presume that such
radial components are extremely weak compared to the transverse
components on large scales.
%In reality, the anisotropy of the field must play a role in
%shocks, but the detailed discussion on the relation between
%magnetic field and shocks is beyond the scope of this paper.
In common with the conventional shock analysis
%\citep[e.g.][]{b7,b8,b9},
(e.g. Zel'dovich \& Raizer 1966, 1967), we use a pair of square
brackets outside each expression enclosed to denote the difference
between the upstream (subscript `1') and downstream (superscript
`2') sides across a MHD shock front. Note that the definitions of
the downstream and upstream sides are in the reference framework
where the shock front is at rest, and the specific entropy
increases from the upstream side to the downstream side.
%We now reduce shock equations (\ref{equ13})$-$(\ref{equ16})
%to self-similar forms for the convenience of analysis.
The sound parameter $k$ in transformation (\ref{equ7}) is related
to the polytropic sound speed and changes across a shock. We
therefore relate upstream $k_1$ and downstream $k_2$ with a ratio
factor $\lambda$ such that
\begin{equation}
k_2=\lambda^2k_1\ ,\quad\qquad h_1=h_2\ ,\quad\qquad x_1=\lambda
x_2\ .\label{equ17}
\end{equation}
%{\bf Stop checking here}
%
The latter two in condition (\ref{equ17}) are for the jump of the
magnetic energy density $<B_t^2>$ and the continuity of shock
radius $r_s$ across the MHD shock front, which means while the
dimensionless shock front $x_{\rm s}$ has different values across
the shock, they correspond to the same shock front radius $r_s$.
With the definition of $h$, $h_1=h_2$ actually means the
transverse magnetic field is proportional to the mass density
across the shock front, consistent with Dyson \& Williams (1997).
%\citet{DW}.
Using $h$ instead of $h_1$ and $h_2$ for the magnetic parameter and
substituting equation (\ref{equ17}) into MHD shock conditions
(\ref{equ13})$-$(\ref{equ16}), we obtain self-similar MHD shock
conditions
\begin{eqnarray}
&&\alpha_1(nx_1-v_1)=\lambda\alpha_2(nx_2-v_2)\ ,\label{equ181}\\
\nonumber
\\ \nonumber
&&\qquad C\alpha_1^{2-n+3nq/2}x_1^{2q}(nx_1-v_1)^q\nonumber\\
&&\qquad\quad +\alpha_1(nx_1-v_1)^2
+\frac{h\alpha_1^2x_1^2}{2}\nonumber\\
&&\qquad\qquad
=\lambda^2\bigg[C\alpha_2^{2-n+3nq/2}x_2^{2q}(nx_2-v_2)^q
\nonumber\\ &&\qquad\qquad\quad
+\alpha_2(nx_2-v_2)^2+\frac{h\alpha_2^2x_2^2}{2}\bigg]\ ,
\label{equ182}\\ \nonumber  &&
(nx_1-v_1)^2+\frac{2\gamma}{(\gamma-1)}C\alpha_1^{1-n+3nq/2}
\nonumber\\
&&\qquad\times x_1^{2q}(nx_1-v_1)^q+2h\alpha_1x_1^2
\nonumber\\
&&\qquad\quad
=\lambda^2\bigg[(nx_2-v_2)^2+\frac{2\gamma}{(\gamma-1)}C
\alpha_2^{1-n+3nq/2}\nonumber\\
&&\qquad\qquad \times x_2^{2q}(nx_2-v_2)^q+2h\alpha_2x_2^2\bigg]\
\label{equ183}
\end{eqnarray}
in terms of the dimensionless reduced variables. Once we have
$(x_2,\ \alpha_2,\ v_2)$ on the downstream side of a shock, we can
determine $(x_1,\ \alpha_1,\ v_1)$ on the upstream side using MHD
shock conditions $(\ref{equ181})-(\ref{equ183})$ (Wang \& Lou
2008) or vice versa. In cases of $q=2/3$, there are only two
independent relations among (\ref{equ181})$-$(\ref{equ183}) (Lou
\& Cao 2008); therefore one may choose parameter $\lambda>0$
fairly arbitrarily, e.g. $\lambda=1$.

\end{document}